\documentstyle[epsf,aps,twocolumn]{revtex}
\draft

\begin{document}

\title{Recoil and momentum diffusion of an atom close to a vacuum-dielectric
interface}
\author{Carsten Henkel\thanks{%
Present address: Institut f\"{u}r Physik,
Universit\"{a}t Potsdam, Am Neuen Palais 10, D-14469~Potsdam, Germany} and
Jean-Yves Courtois\thanks{%
Present address: Thomson-CSF Optronique, Direction Scientifique,
Rue Guynemer, B.P.~55, 78283~Guyancourt, France.}}
\address{Institut d'Optique,\\
Unit\'e de recherche no.14 associ\'ee au CNRS,\\
B.~P.~147, F-91403~Orsay CEDEX, France}
\date{2 February 1998}
\maketitle

\vspace*{-5cm}
\noindent {\large submitted to {\em European Physical Journal D}} 
\vspace*{5cm}

\begin{abstract}
We derive the quantum-mechanical master equation (generalized optical Bloch
equation) for an atom in the vicinity of a flat dielectric surface. This
equation gives access to the semiclassical radiation pressure force and the
atomic momentum diffusion tensor, that are expressed in terms of the vacuum
field correlation function (electromagnetic field susceptibility). It is 
demonstrated that the atomic center-of-mass motion provides a nonlocal
probe of the electromagnetic vacuum fluctuations. We show
in particular that in a circularly polarized evanescent wave, the radiation
pressure force experienced by the atoms is not colinear with the
evanescent wave's propagation vector. 
In a linearly polarized evanescent wave, the recoil per fluorescence 
cycle leads to a net magnetization for a $J_{g}=1/2$ ground state atom.
\end{abstract}


\section{Introduction}

When an atom is absorbing or emitting light, its center-of-mass is subject
to photon recoil. This phenomenon, already pointed out by Einstein in the
early years of the century \cite{Einstein17}, is the core ingredient of 
the atomic motion manipulation techniques that have attracted  much
attention during the last 20 years~\cite{Schawlow75,Cook80,Gordon80,Cohen85a}%
. Quite paradoxically, however, this basic ingredient embodies its own
limit, as it also provides crucial limiting factors to the performances of
these techniques. For example, the minimum attainable temperatures in laser
cooling are generally limited by the random fluctuations in the momenta
exchanged between photons and atoms, that give rise to atomic momentum
diffusion \cite{Einstein17,Schawlow75,Cook80,Gordon80}; also in the most
promising field of atom optics~\cite{Adams94}, spontaneous emission
represents a lethal threat to the coherence of the de Broglie waves because
the atomic wave vector acquires an indeterminacy due to the random photon
recoil \cite{Mlynek94a}.

In this paper, we study the recoil effects due to
spontaneous emission in the vicinity of a vacuum--dielectric
interface for an atom being reflected from an evanescent
wave. This atom-optical device is one of the most studied
realizations of a coherent mirror for atomic de Broglie waves
\cite{Cook82,Balykin87,Chu90,Cohen93b,Aspect96a}. 
Being based upon the interaction between an atom
and an evanescent laser wave propagating along the surface
of a dielectric prism, it already allowed for detailed
experimental investigations of fluorescence rates 
\cite{Mlynek94b,Ertmer94},
single optical pumping cycles \cite{Ovchinnikov95,Dalibard96c} 
and ground-state energy shifts \cite{Landragin96a}
at distances of order $\lambdabar = c / \omega_A$
from the interface ($\omega_A$ is the atomic transition
frequency). Spontaneous emission in the evanescent wave is
also used in reflection cooling techniques 
\cite{Ovchinnikov95,Dalibard96c,Wilkens97b} 
that have been proposed for radiative atom traps in the 
vicinity of surfaces 
\cite{Soeding95,Dalibard96b,Dowling97,Chevrollier97,Pfauetal,Mlynek97f,%
Ovchinnikov97b}. 
From these experiments it
has become apparent that a proper description of the
fluorescence rates and the energy levels has to take
into account the distortion of the electromagnetic field
due to the presence of the dielectric. A precise theory
of the atom--light interaction in the evanescent wave mirror
hence touches upon the field of cavity QED~\cite{Haroche92}
and might even provide a model system for one of this field's
paradigms: that the radiative properties of an atom 
are determined by the local properties of the electromagnetic
field at the atom's position. In fact, as far as the `internal'
atomic dynamics (spontaneous emission rates and frequency shifts) 
is concerned, this problem has already been 
studied intensively, starting from the work of Drexhage~\cite{Drexhage74}
and Chance {\em et al.}\ \cite{Chance78}
and covering a variety of geometries and materials (dielectric
or metallic)~\cite
{Agarwal75d,Cook87,Hinds91,Yoshida93,Haroche94,Ito94,Kim94,Jhe95,%
Polman95,Letokhov96a,Girard95,Rahmani97}.
In a recent paper, Courtois {\em et al.}\ \cite{Courtois96}
calculated the cavity QED modifications to the optical Bloch
equations that govern the relaxation processes of a multilevel
atom close to the vacuum--dielectric interface. This study
showed that the relaxation due to spontaneous emission is
determined by the radiative damping rates of classical dipoles
located near the interface. The paper was limited, however,
to atoms at rest: only the internal dynamics was treated,
whereas the `external dynamics' (recoil of the atomic 
center-of-mass) had still to be included. 
The present paper intends to fill this gap: we derive the
so-called generalized optical Bloch equations that describe
both internal and external dynamics of an atom in the vicinity
of the vacuum--dielectric interface. We actually start from a
more general perspective and determine, for a generic cavity
geometry, the master equation for the atomic density matrix,
including the center-of-mass degrees of freedom. 
The modifications of the electromagnetic vacuum field
appear in this equation through the field correlation
function, taken at two spatially separated positions. 
It hence turns out that the atomic external degrees of freedom
constitute a nonlocal probe of the spatial correlations of
the electromagnetic cavity field.
We then follow the general procedure outlined by Dalibard
and Cohen-Tannoudji \cite{Cohen85a}
and use the Wigner representation to
express the spatial dependence of the atomic density matrix
in terms of a phase-space quasi-distribution. 
In the semiclassical limit, the atomic Wigner function evolves 
according to
a Fokker--Planck equation where appear the radiation pressure
force and the momentum diffusion, and these quantities involve
spatial derivatives of the field correlation function.
In free space, this is simply a reformulation of the random
momentum exchanges between the atom, the laser field, and the
vacuum field \cite{Berman96}. 
In a cavity-type geometry, on the other hand,
photons do not rapidly escape to infinity, and the spatial
structure of the field modes becomes important for
the atomic recoil.

We illustrate the capabilities of the Bloch--Fokker--Planck 
equation derived in this paper by focusing on 
spontaneous emission in the evanescent wave mirror.
In particular, some unusual properties of the radiation pressure 
force above the dielectric interface are displayed.
As a first example, we show that it
differs from the naive estimate based upon the
phase gradient of the evanescent driving field and the spontaneous emission
rate. More explicitly, if the evanescent wave is circularly polarized, the
radiation pressure force points into a different direction than the (real
part of the) evanescent field's wave vector. 
This is due to the fact that the vacuum--dielectric interface
partially reflects the electromagnetic field and hence modifies
the spatial structure of the vacuum fluctuations. 
From the viewpoint of radiation reaction,
the correction to the radiation pressure corresponds to the force
exerted by the atom's dipole field that is backreflected 
from the interface.
As a second example, we study the optical pumping of a $J_{g}=1/2$ atom 
in the vicinity of the dielectric. The reduced symmetry of the
electromagnetic vacuum field implies that the average recoil 
per optical pumping cycle differs between the two Zeeman sublevels, 
even in a linearly polarized evanescent field. 
As a consequence, an initially unpolarized atomic ensemble
splits into two spin components with different average momentum after 
a pumping cycle. For some velocity classes the sublevel
populations then have become imbalanced, and the atomic ensemble
shows what may be called a `recoil-induced magnetization'.

The theory outlined in this paper thus improves previous `heuristic'
approaches to atomic recoil in evanescent waves \cite
{Mlynek94b,Imbert76,Walls94e}, that assume the atomic
fluorescence to be distributed according to the free-space dipole 
radiation pattern. 
Our results are also relevant for radiative
atom traps in the vicinity of material surfaces \cite
{Soeding95,Dalibard96b,Dowling97,Chevrollier97,Pfauetal,Mlynek97f,%
Ovchinnikov97b}, 
where momentum diffusion due to spontaneous emission 
may be one of the limiting factors for the temperature. 
While current atomic mirror experiments are typically limited 
to the transient regime, such traps would allow to study the 
radiation pressure force in evanescent waves in the long-time limit 
(steady state). 
From a more general perspective, the framework presented here
may also be used to predict the center-of-mass motion of cold atoms 
in a high-finesse optical cavity with its electromagnetic field modes 
being confined both in real and frequency space.  
It is an interesting result for the domain of cavity QED
that the external motion of atoms in a cavity 
provides a nonlocal probe of the cavity field correlation function, 
in opposition to internal radiative properties 
that are determined by the field correlations at the same point. 
The examples we develop demonstrate 
that this direction of cavity QED may be investigated with current experiments.

\medskip

The paper is organized as follows: in Sec.~II, we present the generalized
optical Bloch equations including a nontrivial correlation function for the
electromagnetic vacuum field. We focus on atoms driven by a
monochromatic field in the low-saturation, large-detuning limit. Eliminating
adiabatically the excited state, the Bloch equations reduce to the optical
pumping equation involving only the ground state density matrix. Passing to
the Wigner representation, these equations take the form of a Fokker--Planck
equation in the semiclassical limit. We display general expressions for the
radiation pressure force and the momentum diffusion tensor that apply to any
Zeeman degeneracy. The conditions of validity for our approach are
summarized. In Sec.~III, the example of the evanescent wave mirror allows us
to illustrate the general theory. We discuss the reduced symmetry of the
electromagnetic field correlations in the vacuum above a flat dielectric
surface and recover, in the absence of recoil, the well-known fluorescence
rates for this geometry. Specializing to a scalar ground state, {\em i.e.},
a $J_{g}=0\rightarrow J_{e}=1$ atom, we study the influence of the
evanescent wave's polarization on the radiation pressure force and the
momentum diffusion tensor. We then consider a $J_{g}=1/2$ ground state atom
and examine optical pumping in the evanescent wave. 
The appendixes contain several technical results that are used in
the text.

\section{Generalized optical Bloch equations}
\label{s:GOBE}

The internal and external dynamics of an atom interacting with a laser field
are conveniently characterized by a master equation for its density matrix
(generalized optical Bloch equation `G.O.B.E.'). This section is devoted to
the derivation of such an equation in the particular case of a multilevel
atom located close to the interface between vacuum and a dielectric
medium.

\subsection{General}

To begin, we identify the general features of the G.O.B.E. at an interface.
In {\em free space\/}, the master equation describing the interaction of a
single multilevel atom with a monochromatic laser field is well-known \cite
{Cohen77,CastinT}. Basically, its derivation proceeds in two
steps. In the first step, one considers the evolution equation for the
total density matrix of the system constituted by the atom and the
electromagnetic field. In the framework of nonrelativistic quantum
electrodynamics and in the electric dipole approximation, this equation
relies upon the atom-field Hamiltonian 
\begin{equation}
H=H_{0}+H_{R}+V_{AL}+V_{AR}  \label{Htot}
\end{equation}
The first term on the right-hand side of Eq.(\ref{Htot}) is the atomic
Hamiltonian accounting for the internal energy energy of the {\em bare\/} atom
and for its kinetic energy: 
\begin{equation}
H_{0}=\frac{{\bf P}^{2}}{2M}+\frac{\hbar \omega _{0}}{2}(P_{e}-P_{g})
\label{Ha}
\end{equation}
where ${\bf P}$ is the atomic momentum operator, $M$ is the atomic mass, and 
$P_{g}$ and $P_{e}$ are the projection operators on the ground and excited
states, respectively; the second term is the free Hamiltonian of the
Coulomb-gauge quantized electromagnetic field;\ $V_{AL}$ is the
time-dependent, purely atomic Hamiltonian 
\begin{equation}
V_{AL}=-{\bf D}.{\cal \vec{E}}_{L}\left( {\bf R},t\right)  \label{Vav}
\end{equation}
that describes the interaction of the atomic dipole ${\bf D}$ with the
laser field assumed to be in a coherent state and therefore described by a 
{\em classical\/} function ${\cal \vec{E}}_{L}\left( {\bf r},t\right) $; and
the last term, 
\begin{equation}
V_{AR}=-{\bf D}.{\bf E}({\bf R})  \label{Var}
\end{equation}
represents the coupling between the atom and the reservoir associated with
the vacuum quantum field ${\bf E(R})$. We note that in Eq.(\ref{Htot}),
both fields ${\cal \vec{E}}_{L}\left( {\bf R},t\right) $ and ${\bf E(R})$
are evaluated at the location of the atom (${\bf R}$: atomic center-of-mass
position operator). In the second step, the master equation for the atomic
density matrix $\rho $ is obtained by applying second order perturbation
theory to the atom-reservoir interaction, and by tracing away the degrees of
freedom associated with the reservoir. This yields a dynamical evolution
equation where the influence of the reservoir is manifest through two
contributions. The first, associated with an effective Hamiltonian,
describes the energy shifts undergone by the atomic levels as a result of
their coupling to the vacuum field (Lamb-shifts). These shifts are
traditionally assimilated in the definition of $H_{0}$, yielding the actual
atomic Hamiltonian $H_{A,\infty }$. The second contribution, $\dot{\rho}%
_{relax,\infty }$, represents the dissipation of the atomic system due to
its coupling with the reservoir (spontaneous emission). Finally, the
free-space time evolution of the atomic density matrix takes the form 
\begin{equation}
\dot{\rho}={\cal L}_{\infty }\,\rho  \label{equaBlochfs}
\end{equation}
\begin{equation}
{\cal L}_{\infty }\,\rho =\frac{1}{i\hbar }\left[ H_{A,\infty }+V_{AL},\rho
\right] +\dot{\rho}_{relax,\infty }  \label{Liouvillian}
\end{equation}
where we have introduced the free-space Liouville operator ${\cal L}_{\infty
}$\thinspace .

We now consider an atom located in the vicinity of a vacuum-dielectric
interface. What are the modifications of the master equation (\ref
{equaBlochfs}) induced by the lower-lying dielectric medium? First, because
of the new boundary conditions, the modes of ${\cal \vec{E}}_{L}\left( {\bf r%
},t\right) $ and of the quantized electromagnetic vacuum field are altered
and may become evanescent. It is clear that this does not affect the
operators $H_{0}$, $H_{R}$, and $V_{AL}$ which keep the same form as in the
free-space case. In contrast, the structure of the reservoir becomes
modified. The contributions of $V_{AR}$ to the atom dynamics (energy level
shifts and spontaneous emission rates) are therefore expected to be
different from the free-space situation. Moreover, as a result of the
instantaneous Coulomb interaction between the atomic and dielectric charges,
one expects a supplementary electrostatic contribution $H_{es}$ to the
energy level shifts. $H_{es}$ corresponds to the London-Van der Waals
interaction of the instantaneous atomic dipole with its image in the
dielectric medium (higher multipoles can be neglected provided the atomic
radius is much less than the distance between the atom and the dielectric
surface). Finally, denoting by $\Delta H_{A}$ and $\dot{\rho}_{relax,int}$
the modifications of the Hamiltonian and dissipative parts of the atomic
density matrix evolution due to the interface, one obtains the general form
of the G.O.B.E. in the presence of the dielectric medium 
\begin{equation}
\dot{\rho}={\cal L}_{\infty }\,\rho +{\cal L}_{int}\,\rho
\label{equaBlochgen}
\end{equation}
where 
\begin{equation}
{\cal L}_{int}\,\rho =\frac{1}{i\hbar }\left[ \Delta H_{A},\rho \right] +%
\dot{\rho}_{relax,int}  \label{modifeqbloch}
\end{equation}
entirely describes the influence of the interface on the atomic dynamics.\
In particular, ${\cal L}_{int}\,\rho $ tends toward zero when the atom is
far from the dielectric surface. The expression of the atomic level shifts
close to a vacuum-dielectric interface have been presented in Ref. \cite
{Courtois96} and will not be discussed any further. We will therefore only
focus on the dissipative contribution to Eq.(\ref{modifeqbloch}).

\subsection{Master equation treatment of spontaneous emission}

\label{mastereq}

We consider the relaxation processes undergone by the atom as a result of
its coupling with the vacuum quantum field. As is well-known, these
processes are conveniently described by a master equation for the atomic
density matrix. In this section, we derive such an equation taking into
account the presence of the lower-lying dielectric medium.

\subsubsection{Atom-quantum field coupling}

As stated above, the coupling between the atom and the quantized
electromagnetic field (which is responsible for spontaneous emission) is
described by the Hamiltonian $V_{AR}=-{\bf D.E(R})$. The atomic dipole
operator ${\bf D}$ changes sign under parity, and therefore has only zero
matrix elements inside the Zeeman degeneracy subspaces of both the ground
and excited states. Furthermore, because $\left| g\right\rangle $ and $%
\left| e\right\rangle $ have well-defined angular momenta, it is possible
following the Wigner-Eckart theorem to write ${\bf D}$ in terms of a
dimensionless, reduced dipole operator ${\bf d}$%
\begin{equation}
{\bf D}={\cal D\,}{\bf d}  \label{dipolereduit}
\end{equation}
whose matrix elements contain the Clebsch-Gordan coefficients associated
with the addition of the angular momenta $1+J_{g}\rightarrow J_{e}$. In Eq.(%
\ref{dipolereduit}), ${\cal D}$ is a real number characterizing the electric
dipole moment amplitude of the atomic transition. We further decompose the
reduced dipole operator as 
\begin{equation}
{\bf d=}P_{e}\,{\bf d\,}P_{g}+P_{g}\,{\bf d\,}P_{e}={\bf d}^{+}+{\bf d}^{-}
\end{equation}
and expand ${\bf d}^{+}$ and ${\bf d}^{-}=\left( {\bf d}^{+}\right)
^{\dagger }$ onto the standard basis $\{{\bf u}_{\pm 1}=\mp \left( {\bf e}%
_{x}\pm \,i{\bf e}_{y}\right) /\sqrt{2},{\bf u}_{0}={\bf e}_{z}\}$ (where $%
{\bf e}_{x,y,z}$ are the unitary vectors associated with the cartesian
coordinate system) 
\begin{equation}
d_{q}^{+}={\bf d}^{+}.{\bf u}_{q}=\left( d_{q}^{-}\right) ^{\dagger }
\end{equation}
The matrix elements of $d_{q}^{+}$ are then given by the simple expression 
\begin{equation}
\left\langle J_{e}\,M_{e}\right| d_{q}^{+}\left| J_{g}\,M_{g}\right\rangle
=\left\langle J_{g}\,1\,M_{g}\,q\right| J_{e}\,M_{e}\rangle  \label{eltdpq}
\end{equation}
where $\left\langle J_{g}\,1\,M_{g}\,q\right| J_{e}\,M_{e}\rangle $ is the
Clebsch-Gordan coefficient connecting the Zeeman sublevels $\left|
J_{g}\,M_{g}\right\rangle $ and $\left| J_{e}\,M_{e}=M_{g}+q\right\rangle $.
Finally, using the rotating-wave approximation, the interaction Hamiltonian $%
V_{AR}$ takes the more explicit form 
\begin{equation}
V_{AR}=-{\cal D}\sum\limits_{q=-1}^{1}\left( d_{q}^{+}{\,}%
E_{q}^{+}+d_{q}^{-}\,E_{q}^{-}\right)  \label{hamrwa}
\end{equation}
where 
\begin{equation}
{\bf E}^{+}=\sum\limits_{q=-1}^{1}E_{q}^{+}\,{\bf u}_{q}=\left( {\bf E}%
^{-}\right) ^{\dagger }
\end{equation}
is the positive-frequency component of the electric field operator.

\subsubsection{Relaxation equation for the atomic density matrix}

The total contribution $\dot{\rho}_{relax}=\dot{\rho}_{relax,\infty }+\dot{%
\rho}_{relax,int}$ of spontaneous emission to the time evolution of the
atomic density matrix can be readily derived from the standard procedure 
\cite{Agarwal75d,Cohen77,CastinT,MandelWolf} outlined in Appendix~\ref
{a:GOBE}, where it is shown that in spite of the quantization of the
center-of-mass motion, $\dot{\rho}_{relax}$\ is of the familiar form, being
a sum of two terms 
\begin{eqnarray}
\left\langle {\bf r}_{1}\right| \dot{\rho}_{relax}\left| {\bf r}%
_{2}\right\rangle &=&-\frac{\Gamma _{\infty }}{2}\left\langle {\bf r}%
_{1}\right| \left\{ C^{i,j}({\bf R},{\bf R})\,d_{i}^{+}\,d_{j}^{-}\,,\rho
\right\} \left| {\bf r}_{2}\right\rangle  \nonumber \\
&&+\Gamma _{\infty }\,C^{i,j}({\bf r}_{2},{\bf r}_{1})\,\,d_{j}^{-}\,\left%
\langle {\bf r}_{1}\right| \rho \left| {\bf r}_{2}\right\rangle \,d_{i}^{+}
\label{eqnbloch}
\label{EQNBLOCH}
\end{eqnarray}
where $\left\{ A,B\right\} =AB+BA$ denotes the anti-commutator between
operators $A$ and $B$, 
\begin{equation}
\Gamma _{\infty }=\frac{{\cal D}^{2}\omega _{0}^{3}}{3\pi \varepsilon
_{0}\,\hbar \,c^{3}}
\end{equation}
is the natural linewidth of the excited state in free space, and where a sum
is to be taken over the $i,j=x,y,z$ indices. The first line of Eq.(\ref
{eqnbloch}) describes the relaxation of the populations and Zeeman
coherences of the excited state and of the optical coherences due to
spontaneous emission. It involves the dimensionless tensor $C^{i,j}({\bf r}%
_{1},{\bf r}_{2})$, proportional to the Fourier transform of the
electromagnetic\ vacuum field correlation function at the atomic transition
frequency $\omega _{0}$ 
\begin{equation}
\Gamma _{\infty }C^{i,j}({\bf r}_{1},{\bf r}_{2})=\frac{{\cal D}^{2}}{\hbar
^{2}}\int\limits_{-\infty }^{\infty }\!d\tau \,e^{i\omega _{0}\tau }\langle
0|E_{i}^{+}({\bf r}_{1},\tau )E_{j}^{-}({\bf r}_{2},0)|0\rangle
\label{eq:def-fn-corr}
\end{equation}
where $\left| 0\right\rangle $ denotes the vacuum state of the field. The
second line in Eq.(\ref{eqnbloch}) describes the feeding of the
ground-state Zeeman sublevels by spontaneous emission, yielding the expected
population conservation relation $%
%TCIMACRO{\func{Tr}}
%BeginExpansion
\mathop{\rm Tr}%
%EndExpansion
\,\left( \dot{\rho}_{relax}\right) =0$.

It is clearly apparent in Eq.(\ref{eqnbloch}) that the effect of the
dielectric medium on the atomic relaxation is entirely described by the
correlation tensor $C^{i,j}({\bf r}_{1},{\bf r}_{2})$ previously derived by
Carnaglia and Mandel \cite{Mandel71} (a useful representation of $C^{i,j}(%
{\bf r}_{1},{\bf r}_{2})$ is given in Appendix~\ref{a:fn-corr}). Finally, we
note that in the case where the atom is infinitely far from the
vacuum-dielectric interface, the correlation tensor $C^{i,j}({\bf r}_{1},%
{\bf r}_{2})$ reduces to its free-space value \cite{CDG1}, and Eq.(\ref
{eqnbloch}) transforms into its well-known expression: 
\begin{eqnarray}
&& \left\langle {\bf r}_{1}\right| \dot{\rho}_{relax,\infty }\left| {\bf r}%
_{2}\right\rangle \nonumber\\
&& = -\frac{\Gamma _{\infty }}{2}\left\langle {\bf r}%
_{1}\right| \left\{ P_{e}\,,\rho \right\} \left| {\bf r}_{2}\right\rangle 
+\Gamma _{\infty }\,\int \frac{d^{2}{\bf n}}{8\pi /3}\,\sum_{\bbox{%
\varepsilon} \bot {\bf n}}
\nonumber \\
&& \quad \left( {\bf d}^{-}.\bbox{\varepsilon }^{*}\right) \,e^{-i%
{\bf k.r}_{1}}\,\left\langle {\bf r}_{1}\right| \rho \left| {\bf r}%
_{2}\right\rangle e^{i{\bf k.r}_{2}}\,\left( {\bf d}^{+}.\bbox{\varepsilon }%
\right)  \label{eqnblochvac}
\end{eqnarray}
where ${\bf n}$ is a unit vector and ${\bf k}$ is defined by ${\bf k=(}%
\omega _{0}{\bf /}c{\bf )n}$.

\subsection{Evolution of the ground state density matrix in the Wigner
representation}

In this section, the G.O.B.E.~(\ref{equaBlochgen}) is transformed into a
Fokker--Planck-type equation for the phase-space distribution function of
the atomic ground state. First, we eliminate adiabatically the optical
coherences $\rho _{eg}=P_{e}\,\rho \,P_{g}$, $\rho _{ge}=P_{g}\,\rho \,P_{e}$
and the excited state density matrix $\rho _{ee}=P_{e}\,\rho \,P_{e}$ by
expressing them in terms of the ground-state density matrix $\rho
_{gg}=P_{g}\,\rho \,P_{g}\equiv \sigma $. This approximation, which holds in
the limit of large laser frequency detunings from resonance and low
saturation of the atomic transition, presents the advantage of reducing the
atomic dynamics to a single Zeeman manifold (optical pumping equation). In a
second step, we Wigner-transform the ground-state density operator $\sigma $%
. In the new representation, $\sigma $ is represented by a $(2J_{g}+1)\times
(2J_{g}+1)$ matrix $W({\bf r,p},t)$, particularly well suited to the
investigation of the semiclassical limit of the atomic motion \cite
{Cook80,Cohen85a}.

\subsubsection{Adiabatic elimination of the excited state}

\label{s:Hamiltonian}

In laser cooling or atom optics experiments, it is customary to operate in
conditions of large laser frequency detuning from resonance and low
saturation of the atomic transition. These conditions allow one to perform
the adiabatic elimination of both the optical coherences $\rho _{eg}$, $\rho
_{ge}$ and the excited state density matrix $\rho _{ee}$. As shown in
Appendix~\ref{a:elimination}, this elimination amounts to the following
substitutions. First, the dipole operators ${\bf d}^{\pm }$ are replaced by 
\begin{equation}
{\bf d}^{-}\mapsto {\bf b}^{-}({\bf R})={\bf d}^{-}\left[ {\bf d}^{+}\cdot 
\bbox{\xi}({\bf R})\right]  \label{eq:def-operateur-b}
\end{equation}
\begin{equation}
{\cal \vec{E}}_{L}({\bf r})={\cal E}_{0}\,\bbox{\xi}({\bf r})
\label{def-xi}
\end{equation}
with ${\bf b}^{\pm }({\bf R})$ being hermitian conjugates. The
non-normalized, dimensionless vector $\bbox{\xi}({\bf r})$ specifies the
laser spatial profile and polarization, while ${\cal E}_{0}$ gives the order
of magnitude of the electric field amplitude.

The second replacement involves the caracteristic timescale of the
ground-state density matrix elements: the spontaneous emission rate $\Gamma
_{\infty }$ is replaced by the typical photon scattering rate $\Gamma
_{\infty }^{\prime }$ 
\begin{equation}
\Gamma _{\infty }\mapsto \Gamma _{\infty }^{\prime }=\Gamma _{\infty }\frac{%
s_{0}}{2}  \label{eq:def-gamma-prime}
\end{equation}
where
\begin{equation}
s_{0}=2\left( \frac{{\cal DE}_{0}}{\hbar \Delta }\right) ^{2} \ll 1
\label{eq:def-saturation}
\end{equation}
is the saturation parameter in the large detuning limit $%
\left| \Delta =\omega _{L}-\omega _{0}\right| \gg \Gamma _{\infty }$.

Finally, using the usual rotating-wave approximation for $V_{AL}$ and by
neglecting the influence of the atomic velocity on laser frequency detuning
(Doppler effect), the G.O.B.E.~(\ref{equaBlochgen}) yields the optical
pumping equation (see Appendix~\ref{a:elimination}): 
\begin{eqnarray}
\langle {\bf r}_{1}|\dot{\sigma}|{\bf r}_{2}\rangle &=&\frac{1}{i\hbar }%
\langle {\bf r}_{1}|\left[ \frac{{\bf P}^{2}}{2M}+H_{eff}({\bf R}),\,\sigma
\right] |{\bf r}_{2}\rangle 
\nonumber\\
&& -\frac{\Gamma _{\infty }^{\prime }}{2}\langle 
{\bf r}_{1}|\left\{ {\cal G}({\bf R}),\sigma \right\} |{\bf r}_{2}\rangle 
\nonumber \\
&&+\,\Gamma _{\infty }^{\prime }\,C^{i,j}({\bf r}_{2},{\bf r}_{1})b_{j}^{-}(%
{\bf r}_{1})\langle {\bf r}_{1}|\sigma |{\bf r}_{2}\rangle b_{i}^{+}({\bf r}%
_{2})  \label{eqnpompage}
\end{eqnarray}
where 
\begin{eqnarray}
H_{eff}({\bf R}) & = & 
P_{g}\,\Delta H_{A}({\bf R})\,P_{g}
\nonumber\\
&&+\hbar \Delta ^{\prime
}\,\left( {\bf d}^{-}\cdot \bbox{\xi}^{*}({\bf R})\right) \left( {\bf d}%
^{+}\cdot \bbox{\xi}({\bf R})\right)  \label{Heff}
\end{eqnarray}
is the effective Hamiltonian accounting for the dielectric-induced energy
level shifts of the ground state (first term on the right-hand side) and the
ground-state light shifts (last term), the order of magnitude of which is 
\begin{equation}
\hbar \Delta ^{\prime }=\hbar \Delta \,\frac{s_{0}}{2}
\end{equation}
We also introduced in (\ref{eqnpompage}) the ground-state operator 
\begin{equation}
{\cal G}({\bf r})=C^{i,j}({\bf r},{\bf r})b_{i}^{+}({\bf r})b_{j}^{-}({\bf r}%
)  \label{opg}
\end{equation}

Besides, to first order in $\Gamma _{\infty }/\Delta $, the connection
between $\sigma $ and the optical coherences and excited-state parts of the
density matrix can be readily expressed in the following form (see Appendix 
\ref{a:elimination}):
\begin{eqnarray}
&& \langle {\bf r}_{1}|\rho _{eg}|{\bf r}_{2}\rangle =\langle {\bf r}_{2}|\rho
_{ge}|{\bf r}_{1}\rangle ^{\dagger } =
\nonumber\\
&& \quad = -\frac{{\cal DE}_{0}}{\hbar \Delta }\left(
\delta ^{i,j}-i\frac{\Gamma _{\infty }}{2\Delta }C^{i,j}({\bf r}_{1},{\bf r}%
_{1})\,\right) 
\times\nonumber\\
&&\quad \quad \times
d_{i}^{+}b_{j}^{-}( {\bf r}_1 )
\,\langle {\bf r}_{1}|\sigma |{\bf r}_{2}\rangle \,e^{-i\omega _{L}t}
\label{eq:coherences}
\end{eqnarray}
\begin{equation}
\langle {\bf r}_{1}|\rho _{ee}|{\bf r}_{2}\rangle =\frac{s_{0}}{2}\,\left( 
{\bf d}^{+}\cdot \bbox{\xi}({\bf r}_{1})\right) \langle {\bf r}_{1}|\sigma |%
{\bf r}_{2}\rangle \left( {\bf d}^{-}\cdot \bbox{\xi}^{*}({\bf r}_{2})\right)
\label{eq:rho-excite}
\end{equation}
where $\delta ^{i,j}$ is the Kronecker symbol. The validity conditions of
the expressions given in this section are detailed in section \ref
{s:validity}.

\subsubsection{Wigner representation of the ground state density matrix}

\label{s:Wigner}

\paragraph{General.}

The Wigner representation of the density matrix provides a particularly
convenient framework for the intuitive understanding of atomic motion in
laser light. This is because beside its intrinsic quasi-probability
distribution character, which often enables to map classical pictures onto
phenomena of quantum nature such as the atomic recoil induced by absorption
or emission of photons, the Wigner representation is the best suited for
taking advantage of the general characteristics of laser-cooled atomic
samples to exhibit momentum widths $\Delta p$ significantly larger than the
photon momentum $\hbar k$, that characterizes the elementary step of the
momentum random walk experienced by the atom as a result from its momentum
exchanges with the laser field.

More quantitatively, the time evolution of the Wigner representation $W({\bf %
r},{\bf p},t)$ of the ground state atomic density matrix
\begin{equation}
W({\bf r},{\bf p},t)=\frac{1}{(2\pi \hbar )^{3}}\int \!d^{3}s\,
\sigma ({\bf r};{\bf s})
\exp {(-}i{\bf p}.{{\bf s}/\hbar )}  \label{eq:def-Wigner}
\end{equation}
\begin{equation}
\sigma ({\bf r};{\bf s})\equiv \left\langle {\bf r}+{\textstyle \frac{1}{2}}%
{\bf s}|\sigma |{\bf r}-{\textstyle \frac{1}{2}}{\bf s}\right\rangle
\end{equation}
can be deduced from the optical pumping equation (\ref{eqnpompage}). One
thus finds after a straightforward calculation: 
\begin{eqnarray}
&& \frac{\partial W}{\partial t}({\bf r},{\bf p},t) = -\frac{{\bf p}}{M}\cdot
\nabla _{{\bf r}}W({\bf r},{\bf p},t)+
\nonumber\\
&& + 
\int \!\frac{d^{3}{\bf s}}{(2\pi \hbar
)^{3}}e^{-i{\bf p}.{\bf s}/\hbar }\times  \nonumber \\
&&\times \bigg( \frac{1}{i\hbar }\left[ H_{eff}({\bf r}+{\textstyle \frac{1}{%
2}}{\bf s})\sigma ({\bf r};{\bf s})-\sigma ({\bf r};{\bf s})H_{eff}({\bf r}-{%
\textstyle \frac{1}{2}}{\bf s})\right]  \nonumber \\
&&- \frac{\Gamma _{\infty }^{\prime }}{2}\left\{ {\cal G}({\bf r}+{%
\textstyle \frac{1}{2}}{\bf s})\sigma ({\bf r};{\bf s})+\sigma ({\bf r};{\bf %
s}){\cal G}({\bf r}-{\textstyle \frac{1}{2}}{\bf s})\right\} 
\nonumber \\
&&+ \Gamma _{\infty }^{\prime }C^{i,j}({\bf r};{\bf s})b_{j}^{-}({\bf r}%
+{\textstyle \frac{1}{2}}{\bf s})\sigma ({\bf r};{\bf s})b_{i}^{+}({\bf r}-{%
\textstyle \frac{1}{2}}{\bf s}) \bigg)  \label{pompwigner}
\end{eqnarray}
where 
\begin{equation}
C^{i,j}({\bf r};{\bf s})\equiv C^{i,j}\left( {\bf r}-{\textstyle \frac{1}{2}}%
{\bf s},{\bf r}+{\textstyle \frac{1}{2}}{\bf s}\right)
\end{equation}
Because the different quantities inside the integral exhibit an ${\bf s}$
dependence, Eq.(\ref{pompwigner}) clearly connects $\partial _{t}W({\bf r},%
{\bf p},t)$ to some others $W({\bf r},{\bf p}+\delta {\bf p},t)$, which is
reminiscent from the recoil of the atom during photon absorption or emission
processes, hence $\left| \delta {\bf p}\right| \approx \hbar k$. Because $%
\hbar k/\Delta p\ll 1$, Eq.(\ref{pompwigner}) can be accurately evaluated
by expanding $W({\bf r},{\bf p}+\delta {\bf p},t)$ up to second order in $%
\delta {\bf p}$ \cite{Cohen85a}. 
A more direct way of implementing this procedure is to
note that $\hbar k/\Delta p\ll 1$ implies that the coherence length of the
atomic ensemble, $\hbar /\Delta p$, is small compared to the optical
wavelength $\lambda =2\pi /k$. This implies that the width in ${\bf s}$ of
the external coherence function $\sigma ({\bf r};{\bf s})$ is small compared
to the scale of variation of the quantities $C^{i,j}({\bf r}-\frac{1}{2}{\bf %
s},{\bf r}+\frac{1}{2}{\bf s})$ and ${\bf b}^{\pm }({\bf r}\pm \frac{1}{2}%
{\bf s})$, which is of the order of $\lambda $. It is therefore possible to
expand directly the integral kernel of Eq.(\ref{pompwigner}) up to second
order in $k{\bf s}$ before evaluating the integral. We now discuss more
quantitatively this procedure in order to identify the influence of the
vacuum-dielectric interface on the atomic dynamics.

\paragraph{Zeroth order: internal atomic dynamics.}

The lowest (zeroth) order in the expansion of the different quantities in
the integration kernel of Eq.(\ref{pompwigner}) amounts to considering the
atoms as point-like particles, {\em i.e.}, to treating the atomic
translational degrees of freedom classically. It is therefore not surprising
to end with the previously-established optical pumping equation \cite
{Courtois96} for a point atom having a constant velocity ${\bf p}/M$ which
yields, after adiabatic elimination of the excited state and optical
coherences: 
\begin{eqnarray}
{\cal O}[(k{\bf s})^{0}]:\quad \left. \frac{\partial W}{\partial t}\right|
_{0} & = & -\frac{{\bf p}}{M}\cdot \nabla _{{\bf r}}W +
\nonumber\\
&& + 
\frac{1}{i\hbar }\left[
H_{eff}({\bf r}),\,W\right] +\left. \dot{W}_{relax}\right| _{0}
\label{eq:OBE}
\end{eqnarray}
where the effective Hamiltonian $H_{eff}$ only impacts on the evolution of
the ground state Zeeman coherences (the space dependence of $H_{eff}$ has no
effect on the atomic motion to zeroth order in $k{\bf s\,}$) and where 
\begin{eqnarray}
\left. \dot{W}_{relax}\right| _{0} & = & 
-\frac{\Gamma _{\infty }^{\prime }\,}{2}%
C^{i,j}({\bf r};{\bf 0})\{b_{i}^{+}({\bf r})\,b_{j}^{-}({\bf r}%
),\,W\} +
\nonumber\\
&& +
\Gamma _{\infty }^{\prime }\,C^{i,j}({\bf r};{\bf 0})\,b_{j}^{-}(%
{\bf r})\,W\,b_{i}^{+}({\bf r})  \label{eq:operateur-pompage}
\end{eqnarray}
accounts for departure from the ground-state through laser absorption (first
term on the right-hand side) and for feeding of the ground-state by
spontaneous emission (second term), the combined action of which yields
optical pumping. Eq.(\ref{eq:operateur-pompage}) shows that the optical
pumping or fluorescence rates are determined by the one-point correlation
tensor $C^{i,j}({\bf r};{\bf 0})=C^{i,j}({\bf r},{\bf r})$. In free space,
one has $C^{i,j}({\bf r};{\bf 0})=\delta ^{i,j}$ so the position dependence
of the fluorescence and optical pumping rates only arises from the driving
field profile $\bbox{\xi}({\bf r})$ 
[{\em cf.}\ Eq.(\ref{eq:def-operateur-b})]. 
On the other hand, close to a vacuum--dielectric interface, 
as will be shown below, one has $C^{i,j}({\bf r};{\bf 0}) = 
\delta ^{i,j}c^{i}(z)$, so an additional cause for a spatially
varying optical pumping rates appears. As already pointed
out in Ref.~\cite{Courtois96}, this phenomenon is directly connected to the
well-known space-dependence of the damping rates of classical oscillating
dipoles close to a vacuum-dielectric interface.

\paragraph{First order: radiative and level shift-induced forces.}

To first order in $k{\bf s}$, where the effect of the atom-field coupling on
the atomic motion enters into play, the ground-state G.O.B.E.'s expansion
takes the form of a Liouville equation uncovering the force ${\bf F}({\bf r})
$ acting on the atom 
\begin{equation}
{\cal O}[(k{\bf s})^{1}]:\quad \left. \frac{\partial W}{\partial t}\right|
_{1}+{\bf F}({\bf r})\cdot \nabla _{{\bf p}}W=0  \label{eq:Liouville}
\end{equation}
The force operator ${\bf F}({\bf r})$ is actually the sum of three terms: 
\begin{equation}
{\bf F}={\bf F}^{(shift)}+{\bf F}^{(dip)}+{\bf F}^{(sp)}
\end{equation}
where 
\begin{equation}
\Big( {\bf F}^{(shift)}+{\bf F}^{(dip)} \Big) \cdot \nabla _{{\bf p}}W=-{%
\textstyle \frac{1}{2}}\left\{ \nabla _{{\bf r}}H_{eff}({\bf r}),\,\nabla _{%
{\bf p}}W\right\} 
\end{equation}
involves the sum of the force associated with the dielectric-induced energy
level shifts (${\bf F}^{(shift)}$) and of the dipole force (${\bf F}^{(dip)}$%
) associated with the ground-state light-shifts (${\bf F}^{(shift)}$ and $%
{\bf F}^{(dip)}$ are associated with the first and second term of Eq.(\ref
{Heff}), respectively), and where 
\begin{equation}
{\bf F}^{(sp)}={\bf F}_{\text{depart}}^{(sp)}+{\bf F}_{\text{feed}}^{(sp)}
\end{equation}
is the radiation pressure force, having two contributions arising from each
term of the right-hand side of Eq.(\ref{eq:operateur-pompage}),
respectively. One finds: 
\begin{equation}
{\bf F}_{\text{feed}}^{(sp)}({\bf r})\cdot \nabla _{{\bf p}}W=-i\hbar \Gamma
_{\infty }^{\prime }\nabla _{{\bf s}}{\cal A}({\bf r};{\bf 0})\cdot \nabla _{%
{\bf p}}W  \label{eq:feed-force-general}
\end{equation}
where ${\cal A}({\bf r};{\bf s})$ is an operator involving the field
correlation function at two points separated by ${\bf s}$, defined as 
\begin{equation}
{\cal A}({\bf r};{\bf s})\,W\equiv C^{i,j}({\bf r};{\bf s})\,b_{j}^{-}({\bf r%
}+{\textstyle \frac{1}{2}}{\bf s})\,W\,b_{i}^{+}({\bf r}-{\textstyle \frac{1%
}{2}}{\bf s})  \label{eq:definition-A}
\end{equation}
and
\begin{equation}
{\bf F}_{\text{depart}}^{(sp)}({\bf r})\cdot \nabla _{{\bf p}}W=\frac{i}{4}%
\hbar \Gamma _{\infty }^{\prime }\left[ \nabla _{{\bf r}}{\cal G}({\bf r}%
),\nabla _{{\bf p}}W\right]  \label{eq:depart-force-general}
\end{equation}

In order to make the physical content of ${\bf F}_{\text{feed}}^{(sp)}$ more
transparent, let us consider the simple situation where the driving laser
field is a plane wave of wavevector ${\bf q}$, in which case the operators $%
{\bf b}^{\pm }$ take the simple form 
\begin{equation}
{\bf b}^{\pm }({\bf r})={\bf \beta }_{0}^{\pm }\,e^{\mp i{\bf q.r}}
\end{equation}
where ${\bf \beta }_{0}^{\pm }$ are space-independent operators. A
straightforward calculation then yields
\begin{eqnarray}
{\bf F}_{\text{feed}}^{(sp)}({\bf r}).\nabla _{{\bf p}}W & = &
\hbar \Gamma_{\infty }^{\prime }\left( {\bf q\,}C^{i,j}({\bf r};{\bf 0})
\right. \nonumber\\
&& \left.
- i\nabla _{{\bf s}}C^{i,j}({\bf r};{\bf 0})\right) 
\beta _{0j}^{-}.\nabla _{{\bf p}}W\,\beta_{0i}^{+}  
\label{eq:F-sp-feed-o-plane}
\end{eqnarray}
Eq.(\ref{eq:F-sp-feed-o-plane}) shows that ${\bf F}_{\text{feed}}^{(sp)}$
results from two contributions, the physical significance of which can be
deduced by referring to the well-known free space situation:
the first term in parentheses describes the atomic recoil 
$\hbar {\bf q}$ due to the absorption of the driving plane wave photons,
while the second is associated with the atomic recoil during spontaneous
emission. One can thus consider that the presence of the dielectric medium
affects the {\em quantitative\/} value of ${\bf F}_{\text{feed}}^{(sp)}$
through the modification of $C^{i,j}({\bf r};{\bf 0})$, but that it remains 
{\em qualitatively\/} analogous to the free space situation.

The situation is quite different for ${\bf F}_{\text{depart}}^{(sp)}$.
Considering Eqs.~(\ref{eq:depart-force-general}) and (\ref{opg}), one can
indeed note that whereas in free space, the contribution to ${\bf F}_{\text{%
depart}}^{(sp)}$ only arises from the space dependence of the laser field, a 
{\em supplementary\/} contribution shows up in the vicinity of a
vacuum-dielectric interface due to the space-dependence of the one-point
correlation tensor $C^{i,j}({\bf r},{\bf r})$. In the preceding case of a
plane wave driving field, whereas ${\bf F}_{\text{depart}}^{(sp)}$ reduces
zero in free space, one thus obtains a purely dielectric-induced
contribution (cancelling in free space) of the form: 
\begin{equation}
{\bf F}_{{\rm depart}}^{(sp)}.\nabla _{{\bf p}}W=\frac{i}{4}\hbar \Gamma
_{\infty }^{\prime }\nabla _{{\bf r}}C^{i,j}({\bf r};{\bf 0}).\left[ \beta
_{0i}^{+}\beta _{0j}^{-},\nabla _{{\bf p}}W\right]
\end{equation}

\paragraph{Second order: momentum diffusion tensor.}

To second order in $k{\bf s}$, we find a Fokker--Planck equation for the
Wigner matrix. Its diffusion term is given by 
\[
{\cal O}[(k{\bf s})^{2}]:\quad \left. \frac{\partial W}{\partial t}\right|
_{2}=D^{k,l}({\bf r})\frac{\partial ^{2}W}{\partial p_{k}\partial p_{l}}
\]
The momentum diffusion tensor 
\begin{equation}
D^{k,l}({\bf r})=D_{\text{depart}}^{k,l}+D_{\text{feed}}^{k,l}
\label{eq:diffusion-general}
\end{equation}
again contains contributions from the departure and feeding terms on the
right-hand side of Eq.(\ref{eq:operateur-pompage}), respectively: 
\begin{equation}
D_{\text{depart}}^{k,l}\frac{\partial ^{2}W}{\partial p_{k}\partial p_{l}}=%
\frac{\hbar ^{2}\Gamma _{\infty }^{\prime }}{16}\left\{ \frac{\partial ^{2}%
{\cal G}}{\partial r_{k}\partial r_{l}},\frac{\partial ^{2}W}{\partial
p_{k}\partial p_{l}}\,\right\}   \label{dpart1}
\end{equation}
\begin{equation}
D_{\text{feed}}^{k,l}=-\,\frac{\hbar ^{2}\Gamma _{\infty }^{\prime }}{2}%
\frac{\partial ^{2}{\cal A}}{\partial s_{k}\partial s_{l}}({\bf r};{\bf 0})
\label{dpart2}
\end{equation}
As it is well-known \cite{Moelmer94}, this tensor results from various
phenomena: randomness of atomic recoil processes due to laser absorption and
stimulated/spontaneous emission, spatial spreading or shrinking of the
atomic wavepacket due to the space variation of the absorption or optical
pumping rates. Because the dielectric medium affects both, absorption and
spontaneous emission processes, one expects modifications of the atomic
momentum diffusion in the vicinity of the vacuum-dielectric interface.

\subsubsection{Validity conditions of the derivations}

\label{s:validity}

To conclude this section, we summarize the validity conditions for our
approach. The mose stringent condition arises from our approximation that
the excited state density matrix adiabatically follows the ground state
density matrix. This means first that the atoms move little on the scale $%
\lambdabar$ during the lifetime $1/\Gamma _{\infty }$ of the excited state,
or equivalently: 
\begin{equation}
\Gamma _{\infty }\gg \frac{k p}{M},\, \frac{ k \Delta p }{ M }
\label{eq:atomes-lents}
\end{equation}
Second, the force $F_e$ acting on the excited state must be sufficiently
small that during the lifetime $1/\Gamma _{\infty }$,
the shift of the atomic momentum is negligible compared to the width
$\Delta p$ of the momentum distribution:
\begin{equation}
\Gamma _{\infty }\gg \frac{F_e}{\Delta p}  
\label{eq:force-petite}
\end{equation}
Under typical experimental conditions (distance $z \sim 1/k$), 
the force $F_e$ is at most of order \cite{Hinds91}
${\cal D}^2 / \varepsilon_0 z^4 \sim \hbar k \Gamma_\infty / (kz)^4
\sim \hbar k \Gamma_\infty$. Condition~(\ref{eq:force-petite})
hence reduces to the semiclassical regime $\Delta p \gg \hbar k$
we shall suppose throughout this article [{\em cf.}\ 
condition~(\ref{eq:cond-semi-cl}) below].
 
The elimination of the optical coherences is governed by a different
condition: the laser detuning $\Delta $ must be larger than any other
frequency scale in the G.O.B.E.\ ({\em cf.}\ Appendix~\ref{a:elimination}) 
\begin{equation}
|\Delta |\gg \Gamma _{\infty },\,\frac{{\cal DE}_{0}}{%
\hbar },\,\frac{k p}{M},\,\frac{ |\Delta H_A| }{ \hbar }
\label{eq:cond-eliminer-coherences}
\end{equation}
This condition implies a small saturation parameter~(\ref{eq:def-operateur-b}%
) and amounts to neglecting the Doppler shift of the laser frequency.
It also allows to discard the dielectric-induced shift of the 
atomic transition frequency compared to the free-space detuning
(see Ref.~\cite{Courtois96} for more details).

Furthermore, since $\omega _{A}\gg |\Delta |$ (near-resonant excitation),
the atomic dynamics is `frozen' at the timescale of the vacuum field
correlation time. This justifies the Markov approximation made in deriving
the G.O.B.E.\ (\ref{eqnbloch}). We note that if the conditions~(\ref
{eq:atomes-lents}, \ref{eq:cond-eliminer-coherences}) are relaxed, one has
to take into account both the ground- and the excited-state manifolds, with
spontaneous emission inducing transitions between them. This picture is
reminiscent of the dressed-state description \cite{Cohen85b} and has been
used, {\em e.g.}, in Refs.\onlinecite{Mlynek94b,Soeding95}.

As pointed out in Subsec.~\ref{s:Wigner} above, the mechanical effects of
spontaneous emission may be described simply in a semiclassical
way if the atomic momentum distribution varies smoothly 
on the scale of the photon momentum 
\begin{equation}
\Delta p\gg \hbar k  \label{eq:cond-semi-cl}
\end{equation}
Combining with condition~(\ref{eq:atomes-lents}), our approach is limited to
transitions with $\Gamma _{\infty }\gg \omega _{\text{recoil}}\equiv \hbar
k^{2}/2M$. It therefore fails for light atoms like He or Li, {\em e.g.},
while it applies for heavier atoms like Na, Ne$^{*}$, Ar$^{*}$, Rb, Cs \ldots

\section{Atomic motion at an evanescent wave mirror}

In this section, we illustrate the capabilities of the approach developed
above by applying it to the motion of an atom in an evanescent wave mirror,
in the vicinity of a vacuum--dielectric interface. 
We first examine the electromagnetic field correlation tensor in this geometry,
with particular emphasis on its symmetry properties. 
We thus recover the well-known atomic damping rates
above the dielectric interface. The radiation pressure force and the
momentum diffusion tensor are then explicitly calculated for a $%
J_{g}=0\rightarrow J_{e}=1$ (scalar) atomic transition. The optical pumping
processes of a $J_{g}=1/2$ atom in the evanescent wave are also investigated.

\subsection{Electromagnetic field above the dielectric}

\subsubsection{Vacuum field correlation tensor}

The field correlation tensor $C^{i,j}({\bf r};{\bf s})$ in the vacuum
half-space above the dielectric has been calculated by Carnaglia and Mandel 
\cite{Mandel71}. As shown in Appendix~\ref{a:fn-corr}, this tensor can be
conveniently written as the sum of the free-space correlation tensor and an
interface-dependent part: 
\begin{equation}
C^{i,j}=C_{\infty }^{i,j}+C_{int}^{i,j}
\end{equation}
The free-space correlations are given explicitly in Eq.(\ref
{eq:decomposer-vide}), although we may deduce most of their properties from
symmetry considerations. First, due to translational invariance, $C_{\infty
}^{i,j}({\bf r;s})\equiv C_{\infty }^{i,j}({\bf s})$ is independent of the
position ${\bf r}$ and only depends on the difference vector ${\bf s}$.\
Second, due to rotational invariance, the correlation tensor at the same
point ({\em i.e.}\ ${\bf s}={\bf 0}$) is proportional to the Kronecker
symbol $\delta ^{i,j}$. Third, for ${\bf s}\ne {\bf 0}$, the tensor may be
decomposed into an isotropic part proportional to $\delta ^{i,j}$, and a
quadrupolar part proportional to $s^{i}s^{j}-\frac{1}{3}{\bf s}^{2}\delta ^{i,j}$.
The coefficients of this decomposition are scalar functions of ${\bf s}^{2}$ \cite
{Berman96}; we give their expansion for small ${\bf s}$ in Eq.(\ref
{eq:decomposer-vide}).

\begin{figure}
\centerline{%
\epsfxsize=0.8\columnwidth \epsfbox[150 480 450 720]{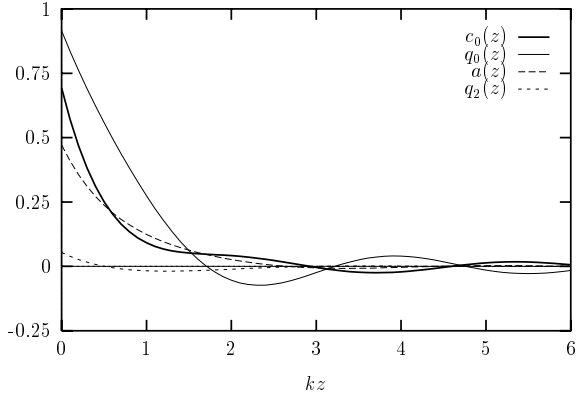}%
}
\caption[fig:c0etc]{%
Dimensionless functions~(\protect\ref{eq:def-c0etc}) 
determining the field correlations as a function of
distance $z$ from the surface (in units of $1/k$). 
\newline
Lateral distance $s_\Vert = 0$, refractive index $n_0 = 1.5$.}
\label{fig:c0etc}
\end{figure}

Let us now apply these symmetry arguments to the interface-dependent part of
the correlation tensor, $C_{int}^{i,j}$. We observe that the translational
and rotational symmetries are broken and reduce to translations parallel to
the interface and rotations around the interface normal, respectively. As a
consequence, we expect $C_{int}^{i,j}({\bf r}_{1},{\bf r}_{2})$ to depend on
the distances $z_{1},\,z_{2}$ of the interface and on the in-plane
difference vector ${\bf s}_{\Vert }={\bf r}_{\Vert ,2}-{\bf r}_{\Vert ,1}$
[the $\Vert $ subscript denotes the translational directions parallel to the
interface ($x,y$ components)]. More precisely, because the 
contributions to $C^{i,j}$ arise from the partial reflection of the field 
at the vacuum--dielectric interface 
and the evanescent waves present in the vicinity of the dielectric 
[{\em cf.}\ Eq.(\ref{eq:corr-int-u-du})], 
$C_{int}^{i,j}$ is expected to depend only on the {\em sum\/} $z_{1}+z_{2}$
of the distances from the interface. Therefore, we may write 
\begin{equation}
C_{int}^{i,j}({\bf r}-{\textstyle \frac{1}{2}}{\bf s},{\bf r}+{\textstyle 
\frac{1}{2}}{\bf s})=C_{int}^{i,j}(z;{\bf s}_{\Vert }).
\label{eq:c-int-dependance}
\end{equation}
As shown in Appendix \ref{a:fn-corr}, this correlation tensor actually
contains four parts having different symmetry properties:
\begin{eqnarray}
&&C_{int}^{i,j}(z;{\bf s}_{\Vert }) = c_{0}(z;{\bf s}_{\Vert }^{2})\,\delta
^{i,j}+q_{0}(z;{\bf s}_{\Vert }^{2})
\Big( \delta ^{z,i}\delta ^{z,j}-{\textstyle \frac{1}{3}}%
\delta ^{i,j}\Big)  \nonumber \\
&&+k\,a_{1}(z;{\bf s}_{\Vert }^{2})
\Big( \delta ^{z,i}\,s_{\Vert }^{j}-s_{\Vert
}^{i}\,\delta ^{z,j}\Big)  \nonumber \\
&&+k^{2}\,q_{2}(z;{\bf s}_{\Vert }^{2})\Big( s_{\Vert }^{i}\,s_{\Vert }^{j}-%
{\textstyle \frac{1}{2}}
{\bf s}_{\Vert }^{2}(\delta ^{i,j}-\delta ^{z,i}\delta ^{z,j})\Big)
\label{cint}
\end{eqnarray}
The first term on the right-hand side of Eq.(\ref{cint}) provides a scalar
contribution, whereas the second has the symmetry of the $Y_{0}^{2}$
spherical harmonic with respect to the interface normal
(quadrupolar part). Note that these two terms entirely
account for the one-point correlation tensor $C_{int}^{i,j}(z;{\bf 0})$,
which is related to the modifications of the natural widths of the excited
state Zeeman sublevels by the interface, as shown below. The third
contribution to Eq.(\ref{cint}), proportional to $\epsilon ^{i,j,k}({\bf s}%
_{\Vert }\times {\bf e}_{z})_{k}$, displays an axial symmetry. Finally, the
fourth term corresponds again to a quadrupolar part with respect to
the in-plane vector ${\bf s}_\Vert$. The scalar weight
functions $c_{0}$, $q_{0}$, $a_{1}$ and $q_{2}$ are given in Apppendix \ref
{a:fn-corr} [Eqs.~(\ref{eq:def-c0etc})] and are plotted in 
Fig.~\ref{fig:c0etc} for ${\bf s}_{\Vert }={\bf 0}$ as a function of $z$. 
It clearly appears on this figure
that the influence of the interface is only significant for distances
smaller than the optical wavelength $\lambda $.

\begin{figure}
\centerline{%
\epsfxsize=0.6\columnwidth \epsfbox[100 00 350 300]{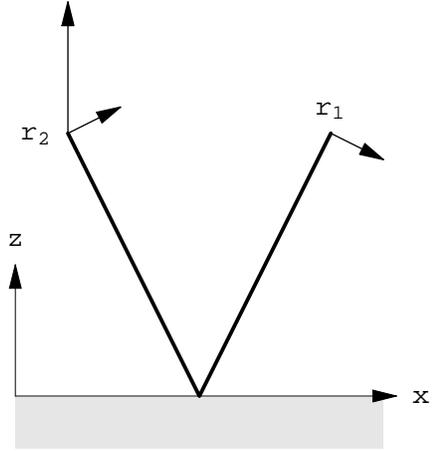}%
}
\caption[fig:dipole-TM]{Illustration of the axial part
of the correlation tensor: electric field created at
${\bf r}_1$ by a dipole located at ${\bf r}_2$ and
oscillating perpendicular to the interface.}
\label{fig:dipole-TM}
\end{figure}

As will be shown in the following, the axial part of the field correlations
is at the origin of nonstandard effects close to the dielectric surface, so
a physical interpretation of this term might be helpful. To this end, we use
the fact that the correlation tensor $C^{i,j}({\bf r}_{1},{\bf r}_{2})$ is
proportional to the electromagnetic field susceptibility, {\em i.e.}\ the
electric field $E_{i}$ created at ${\bf r}_{1}$ by a classical dipole 
located at ${\bf r}_{2}$, oriented along the ${\bf e}_{j}$ axis 
and oscillating at the atomic resonance frequency 
\cite{Agarwal75d,Yoshida93,Girard95,Rahmani97}
({\em cf.}\ also Appendix~\ref{a:susceptibility}). Consider now
a dipole at ${\bf r}_{2}=({\bf 0},z)$, oscillating at the atomic resonance
frequency and oriented 
perpendicular to the interface, and the field it creates at ${\bf r%
}_{1}=({\bf s}_{\Vert },z)$ (see Fig.~\ref{fig:dipole-TM}). If the dipole's
distance $z$ is large compared to the optical wavelength, we may use
geometrical optics to find the rays that reach the observation point ${\bf r}%
_{1}$. As far as $C_{int}^{i,j}$ is concerned, the only ray to consider is
the one that reaches ${\bf r}_{1}$ after one reflection from the interface
(the thick solid line in Fig.~\ref{fig:dipole-TM}). The vertical orientation
of the dipole implies that this ray must be $TM$-polarized. Due to the
finite distance $s_{\Vert }$ parallel to the interface, the reflected
field vector at ${\bf r}_{1}$ 
has a nonzero component parallel to the interface
(actually, parallel to ${\bf s}_{\Vert }$). This construction hence
illustrates how the reflection at the interface creates a correlation
between lateral and perpendicular field components at spatially separated
positions [described by the axial part of the correlation tensor
(\ref{cint})]. 
If the dipole's distance $z$ is not large compared to $\lambdabar$, 
geometrical optics fails and one has to take into account a continuous
distribution of modes that also contains evanescent waves. However, these
modes still have the common feature of being $TM$-polarized and 
therefore also contribute to the axial coefficient 
$a_{1}(z;{\bf s}_{\Vert }^{2})$ of the correlation tensor
[see the Sommerfeld integral~(\ref{eq:def-a})].

\subsubsection{Connection with the damping rates of the excited state Zeeman
sublevels}

As shown in Ref.~\cite{Courtois96}, the atomic internal relaxation processes
associated with spontaneous emission close to a vacuum-dielectric interface
can be entirely described by means of the damping rates $\Gamma _{\Vert }(z)$
and $\Gamma _{\bot }(z)$ of classical oscillating dipoles located at a
distance $z$ above the dielectric medium, polarized parallel or orthogonal
to the interface, respectively, and such that $\Gamma _{\Vert ,\bot
}(z\rightarrow \infty )=\Gamma _{\infty }$. More precisely, one shows that
the contribution $\dot{\rho}_{ee,relax}$ of spontaneous emission to the
evolution of the excited state part of the internal atomic density matrix
for an atom at rest in ${\bf r}$ reads \cite{Courtois96} 
\begin{eqnarray}
&& \dot{\rho}_{ee,relax} = 
\label{rho_ee_relax}\\
&& =
-\frac{1}{2}\left\{ \Gamma _{\Vert
}(z)\,d_{x}^{+}d_{x}^{-}+\Gamma _{\Vert }(z)\,d_{y}^{+}d_{y}^{-}+\Gamma
_{\bot }(z)\,d_{z}^{+}d_{z}^{-},\rho _{ee}\right\}  
\nonumber
\end{eqnarray}
By comparing Eq.(\ref{rho_ee_relax}) with Eq.(\ref{eqnbloch}), which
yields 
\begin{equation}
\dot{\rho}_{ee,relax}=-\frac{\Gamma _{\infty }}{2}\left\{ C^{i,j}({\bf r},%
{\bf r})\,d_{i}^{+}\,d_{j}^{-}\,,\rho _{ee}\right\}
\end{equation}
one can readily deduce that
\begin{mathletters}
\label{eq:taux-modifies}
\begin{eqnarray}
c_{\Vert }(z) &=&C^{x,x}(z;{\bf 0})=C^{y,y}(z;{\bf 0})=\frac{\Gamma _{\Vert
}(z)}{\Gamma _{\infty }} = 
\nonumber\\
& = & 1+c_{0}(z;0)-{\textstyle \frac{1}{3}}q_{0}(z;0) \\
c_{\perp }(z) &=&C^{z,z}(z;{\bf 0})=\frac{\Gamma _{\perp }(z)}{\Gamma
_{\infty }}=
\nonumber\\
& = & 1+c_{0}(z;0)+{\textstyle \frac{2}{3}}q_{0}(z;0)
\end{eqnarray}
As can be checked on the expression of $c_{0}$ and $q_{0}$ [Eqs.~(\ref
{eq:def-c0}) and (\ref{eq:def-q0})], these results are consistent with the
well-known form of $\Gamma _{\Vert }(z)$ and $\Gamma _{\bot }(z)$ \cite
{Chance78,Courtois96}.

\subsubsection{Evanescent driving field}

We focus in this paper on an evanescent driving laser field without
polarization gradient, {\em i.e.}, we consider the field profile 
\end{mathletters}
\begin{equation}
\bbox{\xi}({\bf r})=\bbox{\xi}_{0}\exp {(-\kappa }z{+}iQx{)}
\label{eq:profil-champ}
\end{equation}
where $Q^{2}-\kappa ^{2}=k^{2}$ and $\bbox{\xi}_{0}$ is a constant vector.
This field is created by total internal reflection of a plane laser wave
inside the dielectric (hence, $Q>k$). For the two elementary polarizations $%
TE$ and $TM$ of the incident wave, we have 
\begin{equation}
\bbox{\xi}_{0}^{(TE)}={\bf e}_{y},\qquad \bbox{\xi}_{0}^{(TM)}=\frac{i\kappa
\,{\bf e}_{x}-Q\,{\bf e}_{z}}{k}.  \label{eq:def-f-TE/TM}
\end{equation}
Note that in the $TM$-case, the polarization of the evanescent wave is
elliptic: it approaches a linear polarization in the vicinity of the
critical angle ($\kappa \to 0$) and a circular one ($\sigma ^{-}$ with
respect to the positive $y$-axis) far from the critical angle ($\kappa
\simeq Q$).

\subsection{{$J_{g}=0$}$\to J_{e}=1$ atomic transition}

\label{s:scalaire}

In this subsection we consider the simple situation of a scalar atom
(a $J_{g}=0\to J_{e}=1$ transition) and calculate the radiation 
pressure force and the momentum diffusion tensor. 
The Wigner function of the ground state is now a scalar, and 
the atomic dipole operators ${\bf b}^{\pm }({\bf r})$~[Eq.(\ref
{eq:def-operateur-b})] reduce to $c$-number functions that are simply given
by the electric field profile 
\begin{equation}
{\bf b}^{-}({\bf r})=\bbox{\xi}({\bf r}).  \label{eq:B-scalaire}
\end{equation}
The advantage of such a transition is that it provides a good way to
single out the effect of the interface on the basic atomic external
dynamics, with the minimum complications introduced by the interface-modified
internal atomic dynamics.

\subsubsection{Fluorescence rate}

The atomic ground state reducing to a single level, the only nontrivial
feature of the optical pumping equation (\ref{eq:OBE}) is the total atomic
fluorescence rate $\Gamma ^{\prime }(z)$, that is given by the trace of
either term of the right-hand side of Eq.(\ref{eq:operateur-pompage}). One
thus finds 
\begin{equation}
\Gamma ^{\prime }(z)=\Gamma _{\infty }^{\prime }\left( c_{\Vert }(z)|%
\bbox{\xi}_{0\Vert }|^{2}+c_{\perp }(z)|\xi _{0\bot }|^{2}\right) e^{-2\kappa z}
\label{eq:taux-fluo-scalaire}
\end{equation}
with $\bbox{\xi}_{0}=(\bbox{\xi}_{0\Vert },\xi _{0\bot })$. 
The influence of the interface on the internal ground state dynamics 
hence amounts to a different space-dependence of the broadening of the 
ground state level in addition to the one associated with 
the space dependence of the driving laser field. For comparison,
the parenthesis in Eq.(\ref{eq:taux-fluo-scalaire}) 
is space independent in free space and equal to $|\bbox{\xi}_0|^{2}$. 
It is also interesting to note that the fluorescence rate
(\ref{eq:taux-fluo-scalaire}) involves the
intensity of the `transverse' ($|\xi _{0\bot }|^{2}$) and `longitudinal' 
($|\bbox{\xi}_{0\Vert }|^{2}$) parts of the driving field 
independently, as a result
from the rotational symmetry breaking due to the interface. As a
consequence, the two elementary polarizations of the evanescent wave yield
different spatial variations of the fluorescence rate: 
\begin{mathletters}
\label{eq:taux-TE-TM}
\begin{eqnarray}
\Gamma ^{\prime (TE)}(z) &=&\Gamma _{\infty }^{\prime }\,c_{\Vert
}(z)\,e^{-2\kappa z} \\
\Gamma ^{\prime (TM)}(z) &=&\,\Gamma _{\infty }^{\prime }\,\frac{\kappa
^{2}c_{\Vert }(z)+Q^{2}c_{\perp }(z)}{k^{2}}\,e^{-2\kappa z}
\end{eqnarray}
Hence, depending on the polarization of the driving evanescent wave, the
atomic fluorescence permits to probe different combinations of the
correlation tensor components $c_{\Vert }(z)$ and $c_{\perp }(z)$.

\subsubsection{Radiation pressure force}

\label{s:force-scalaire}

By inserting the field correlation tensor (\ref{eq:decomposer-vide}, \ref
{eq:decomposer-int}) and the evanescent field profile~(\ref{eq:profil-champ}%
) into the general results~(\ref{eq:feed-force-general}, \ref
{eq:depart-force-general}), one readily finds that the departure contribution~(%
\ref{eq:depart-force-general}) to the radiation pressure vanishes for a
scalar atom. After some algebra, the feeding 
term~(\ref{eq:feed-force-general}) yields the following
result for the radiation pressure force 
\end{mathletters}
\begin{eqnarray}
{\bf F}^{(sp)}({\bf r})& = & 
\Gamma ^{\prime }(z)\,\hbar Q\,{\bf e}_{x}+
\nonumber\\
&& + 2\Gamma_{\infty }^{\prime }\,\hbar k\,a_{1}(z;0)
\,\text{Im}\,(\bbox{\xi}_{0\Vert
}\xi _{0\bot }^{*})\,e^{-2\kappa z}  \label{eq:force-scalaire}
\end{eqnarray}
The first term of the force~(\ref{eq:force-scalaire}) corresponds to the
rule-of-the-thumb expression for the radiation pressure: it is the product
of the fluorescence rate~(\ref{eq:taux-fluo-scalaire}) and the photon
momentum $\hbar Q$ carried by the evanescent wave along its propagation
direction. The second term arises from the fact that the actual driving
field consists of the sum of the incoming laser evanescent wave and of the
reflected part of the field radiated by the atomic dipole. Because the
radiation pressure force exerted on a $J_{g}=0\to J_{e}=1$ atom is
proportional to the phase-gradient of the total driving field, it actually
appears as the sum of a contribution proportional to the phase gradient of
the evanescent incoming wave [first term of Eq.(\ref{eq:force-scalaire})] 
{\em and\/} of a term involving the phase gradient of the reflected dipole
field [second term of Eq.(\ref{eq:force-scalaire})]. It is therefore not
surprising to find that the reflected field contribution to the radiation
pressure is proportional to the axial coefficient $a_{1}$ that, as
previously discussed, is directly connected to field reflection processes at
the interface.

\begin{figure}
\centerline{%
\epsfxsize=0.8\columnwidth \epsfbox[120 -20 500 200]{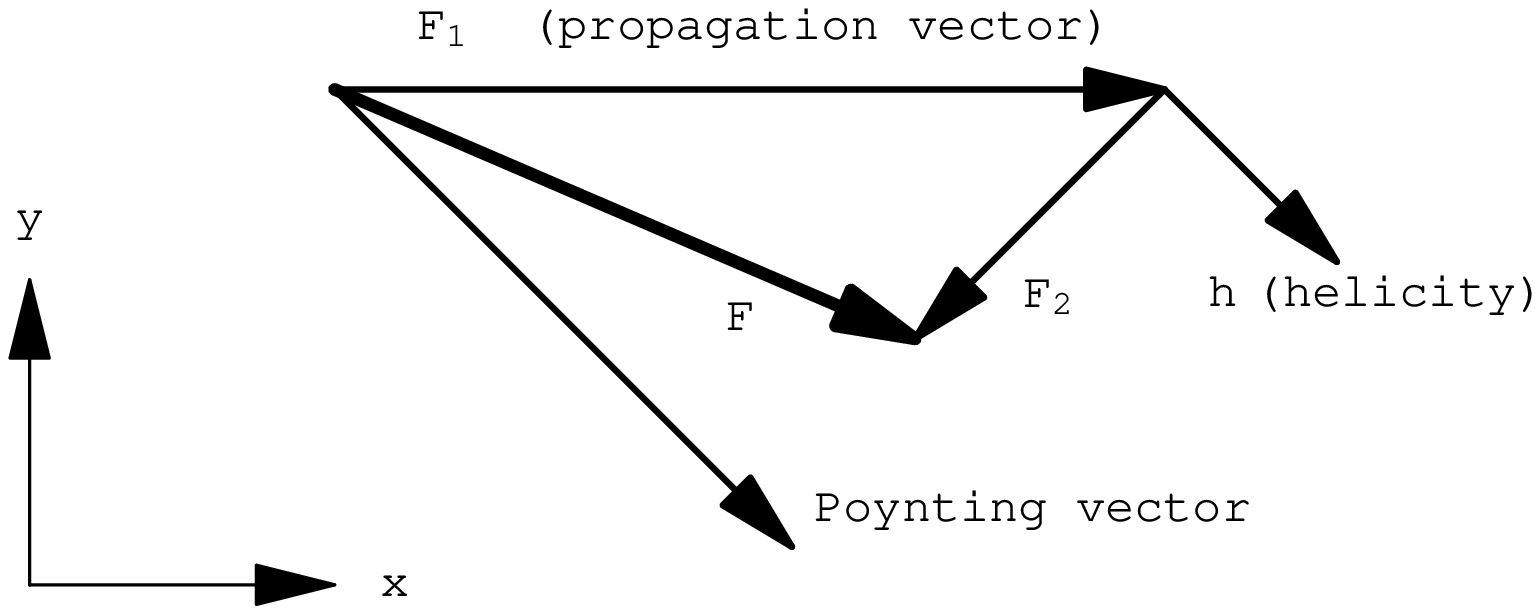}%
}
\caption[fig:polar-sigma]{Illustration of the 
radiation pressure force in a circularly
polarized evanescent wave.
\newline
F$_1$: usual radiation pressure (parallel to driving field's
propagation vector); F$_2$: correction due to partial field
reflection at the dielectric surface; F: full force.
The helicity vector h of the circular polarization and the
Poynting vector of the evanescent wave
(using the standard definition) are also shown.}
\label{fig:polar-sigma}
\end{figure}
\begin{figure}
\centerline{%
\epsfxsize=0.8\columnwidth \epsfbox[150 400 450 650]{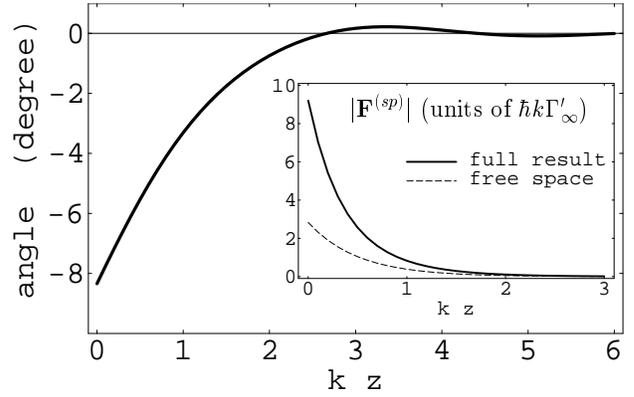}%
}
\caption[fig:force-scalaire]{Direction of the 
radiation pressure force in a circularly polarized 
evanescent wave, as a function of distance from the 
dielectric. The curve shows the angle between the
actual force and the evanescent wave's
propagation vector (the vectors F and F$_1$ of 
Fig.~\ref{fig:polar-sigma}). For these parameters, 
the field's Poynting vector forms an angle of 
$-45^\circ$ with the propagation vector~F$_1$.
The inset gives the
magnitude of the force, using either the exact
result~(\ref{eq:force-sigma}) (full line)
or ignoring both the increased fluorescence rate
and the force correction F$_2$ (dashed line).
\newline
Parameters: refractive index $n_0 = 1.5$, evanescent driving field
with $\kappa = k$, $Q = \sqrt{2}\,k$. The force is plotted is units
of $\hbar k \Gamma'_\infty$, as a function of $kz$.}
\label{fig:force-scalaire}
\end{figure}

For the $TE$ and $TM$ polarizations of the evanescent driving field, this
correction to the radiation pressure force is difficult to observe: indeed,
the vector $\text{Im}\,(\bbox{\xi}_{0\Vert }\xi _{0\bot }^{*})$ vanishes in
the $TE$-case and is parallel to ${\bf e}_{x}$ in the $TM$-case [{\em cf.\/}
Eq.(\ref{eq:def-f-TE/TM})], thus modifying slightly the magnitude of ${\bf F}%
^{(sp)}$. A more prominent modification occurs for a generic combination of $%
TE$ and $TM$ polarizations. The effect is actually maximum for a {\em %
circular\/} polarization of the evanescent wave in a plane perpendicular to
the interface, which can be achieved for 
\begin{equation}
\bbox{\xi}_{0}^{(\sigma)}=\bbox{\xi}_{0}^{(TM)}+i\bbox{\xi}_{0}^{(TE)}.
\label{eq:polar-sigma}
\end{equation}
The field's polarization is then $\sigma ^{+}$ with respect to an axis
parallel to the interface (given by the `helicity' vector ${\bf h}=(2Q/k)[%
{\bf e}_{x}-(\kappa /k){\bf e}_{y}]$, {\em cf.}\ Fig.~\ref{fig:polar-sigma}%
). The total fluorescence rate is given by 
\begin{equation}
\Gamma ^{\prime (\sigma)}(z)
= \Gamma _{\infty }^{\prime }\,\frac{Q^{2}}{k^{2}}%
\,\left( c_{\Vert }(z)+c_{\perp }(z)\right) \,e^{-2\kappa z}
\label{eq:taux-fluo-sigma}
\end{equation}
and the reflected field contribution to the radiation pressure force is
found {\em perpendicular\/} to the helicity ${\bf h}$, yielding 
\begin{eqnarray}
{\bf F}^{(sp)}(z)& = &
\hbar Q \Gamma ^{\prime (\sigma)}(z) {\bf e}_{x}
\nonumber\\
&& - 2\hbar Q \Gamma_{\infty }^{\prime }
\,a_{1}(z;0)\,e^{-2\kappa z}
\left( \frac{\kappa }{k}{\bf e}_{x}+{\bf e}_{y}\right) .  
\label{eq:force-sigma}
\end{eqnarray}
Thus, the radiation pressure force forms a nonzero, $z\,$-dependent angle
with the evanescent wave propagation vector $Q\,{\bf e}_{x}$. This is
represented in Fig.~\ref{fig:force-scalaire} where the angle and magnitude
of the radiation pressure force~(\ref{eq:force-sigma}) are plotted as a
function of the distance $z$ from the dielectric surface.

We finally note that a related effect was discussed by Roosen and Imbert 
\cite{Imbert76}: these authors pointed out that the Poynting vector of an
evanescent wave is not parallel to its propagation vector if the wave is
circularly polarized. If one assumes the radiation pressure force to be
parallel to the Poynting vector of the driving field, which seems tempting
because the Poynting vector represents the local momentum of the field, one
thus finds a result reminding of~(\ref{eq:force-sigma}). In fact, it is
well-known that this assumption is not correct, as can be readily checked on
the simple example of a $J_{g}=0\to J_{e}=1$ atom interacting in free space
with two plane waves 
\begin{equation}
\bbox{\xi}({\bf r})=e^{ikx}{\bf e}_{z}+e^{iky}{\bf e}_{x}
\end{equation}
leading to a Poynting vector 
\begin{equation}
{\bf \Pi }\propto {\bf e}_{x}+{\bf e}_{y}-\cos \left[ k(x-y)\right] \,{\bf e}%
_{z}
\end{equation}
whereas the radiation pressure force is clearly oriented along ${\bf e}_{x}+%
{\bf e}_{y}$. This problem can be readily solved by noticing that 
the Poynting vector ${\bf \Pi }$ is only defined up to a curl,
as shown by the continuity equation 
\begin{equation}
{\bf \nabla .\Pi +\partial }_{t}\varepsilon =0
\end{equation}
where $\varepsilon $ is the electromagnetic energy density. It is thus
straightforward to show that it is always possible to define a ``novel''
Poynting vector satisfying the continuity equation and being parallel to the
radiation pressure force.

\subsubsection{Momentum diffusion tensor}

The momentum diffusion tensor can be readily obtained from Eq.(\ref
{eq:diffusion-general}) using the field correlations~(\ref
{eq:decomposer-vide}, \ref{eq:decomposer-int}) and replacing the dipole
operators ${\bf b}^{\pm }({\bf r})$ by the evanescent field profile
according to Eq.(\ref{eq:profil-champ}). One thus finds
\begin{equation}
D^{k,l}( z ) = 
\frac{\hbar ^{2}}{8}\frac{\partial ^{2}\Gamma ^{\prime }(z)}{%
\partial z^{2}}\,\delta ^{k,z}\delta ^{l,z}\,-\,\frac{\hbar ^{2}\Gamma
_{\infty }^{\prime }}{2}\frac{\partial ^{2}{\cal A}}{\partial s_{k}\partial
s_{l}}(z;{\bf 0})  \label{diff0}
\end{equation}
where
\begin{equation}
{\cal A}(z;{\bf s})=C^{i,j}(z;{\bf s})\,\xi _{0i}^{*}\,\xi _{0j}\,\exp
(-2\kappa z+iQ\,{\bf s.e}_{x})
\end{equation}
We now discuss the physical significance of Eq.(\ref{diff0}) in more
details, with an emphasis on the influence of the interface on momentum
diffusion. We start by considering the first term on the right-hand side of
Eq.(\ref{diff0}), that corresponds to $D_{\text{depart}}^{k,l}$ [see Eq.(%
\ref{dpart1})]. In order to single out the interface contribution, it is
convenient to write the total fluorescence rate in the form
\begin{equation}
\Gamma ^{\prime }(z)=\Gamma _{\infty }^{\prime }\left| \bbox{\xi}_{0}\right|
^{2}\,e^{-2\kappa z}+\Gamma _{int}^{\prime }(z)
\end{equation}
yielding two contributions to $D_{\text{depart}}^{z,z}$%
\begin{equation}
D_{\text{depart}}^{z,z}( z )
= \frac{\hbar ^{2}\kappa ^{2}\Gamma _{\infty }^{\prime
}\left| \bbox{\xi}_{0}\right| ^{2}}{2}\,e^{-2\kappa z}+\frac{\hbar ^{2}}{8}%
\frac{\partial ^{2}\Gamma _{int}^{\prime }(z)}{\partial z^{2}}
\label{ddepart}
\end{equation}
The first term on the right-hand side of Eq.(\ref{ddepart}) does not
involve any surface-induced effects (apart from the existence of the
evanescent driving field) and is associated with the shrinking of the atomic
spatial coherence (hence a broadening in momentum space) due to the
non-uniform, exponential photon absorption probability. This corresponds to
the intuitive fact that the coherence of an atomic wavepacket incident on an
evanescent wave mirror will be destroyed more efficiently by
absorption-spontaneous emission cycles close to the interface than far away
as a result of the inhomogeneous laser intensity. The second term of Eq.(%
\ref{ddepart}) is a correction to $D_{\text{depart}}^{z,z}$
arising from the modification of the fluorescence rate by the interface,
that is responsible for an additional spatial modulation of the photon
absorption probability, hence an additional cause for momentum diffusion.
As is well-known, the dipole damping rates $\Gamma _{\Vert }(z)$ and $%
\Gamma _{\bot }(z)$ display an exponential dependence close to the interface,
and then tend toward their asymptotic value $\Gamma _{\infty }$ with some
damped oscillations \cite{Courtois96}. $\Gamma _{int}^{\prime }(z)$, and
consequently $D_{\text{depart}}^{z,z}$, are therefore expected to exhibit
the same kind of behaviour [see Eqs.~(\ref{eq:taux-fluo-scalaire}) and (\ref
{ddepart})].

Consider now the second term on the right-hand side of Eq.(\ref{diff0}),
corresponding to $D_{\text{feed}}^{k,l}$ [see Eq.(\ref{dpart2})]. Again,
the influence of the interface can be sorted out by expressing ${\cal A}$ in
the form
\begin{eqnarray}
{\cal A}(z;{\bf s)} &=&{\cal A}_{\infty }(z;{\bf s})
+ {\cal A}_{int}(z;{\bf s}_{\Vert }) \\
{\cal A}_{\infty }(z;{\bf s}) &=&C_{\infty }^{i,j}({\bf s})\,\xi
_{0i}^{*}\,\xi _{0j}\,\exp (-2\kappa z+iQ\,{\bf s.e}_{x}) \\
{\cal A}_{int}(z;{\bf s}_{\Vert }) &=&C_{int}^{i,j}(z;{\bf s}_{\Vert })\,\xi
_{0i}^{*}\,\xi _{0j}\,\exp (-2\kappa z+iQ\,{\bf s.e}_{x})
\end{eqnarray}
Because ${\cal A}_{\infty }$ only involves the free-space correlation tensor 
$C_{\infty }^{i,j}$, its contribution to $D_{\text{feed}}^{k,l}$ is
analogous to that usually encountered in free space for any driving laser
field, {\em i.e.}, it accounts for the random walk of the atoms in momentum
space due to their recoil during spontaneous emission of photons, {\em the
spontaneous emission diagram being assumed as in free space}.
The contribution of ${\cal A}_{int}$ to momentum diffusion,
involving the complex nonlocal correlations between the vacuum field
components induced by the interface, exhibits an interesting feature.
Because ${\cal A}_{int}$ is independent of the $z$-component of the relative
position ${\bf s}$ (because such is $C_{int}^{i,j}$), the interface-induced
modifications of momentum diffusion due to changes in spontaneous emission
in the vicinity of the dielectric medium will only take place {\em parallel\/}
to the interface, a result that is not obvious when considering the
important modifications of the spontaneous emission diagrams due to the
interface \cite{Courtois96}.

\begin{figure}
\centerline{%
\epsfxsize=0.8\columnwidth \epsfbox[150 250 450 720]{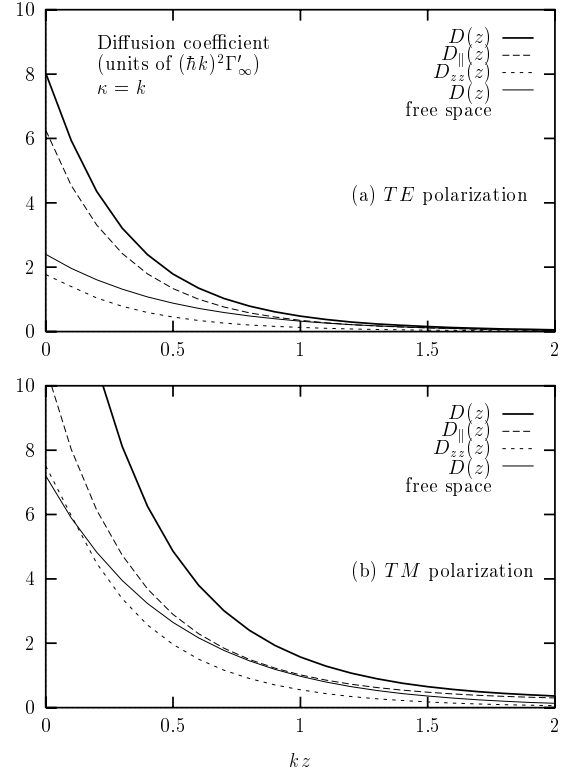}%
}
\caption[fig:trace-D]{Momentum diffusion coefficient
for a scalar atom driven by a
$TE$ [(a)] and a $TM$ [(b)] polarized evanescent wave.
\par
Thick solid line: trace $D(z)$ (\ref{eq:trace-D}) of the 
diffusion tensor;
dashed line: lateral diffusion coefficient 
$D_\Vert(z) = D_{xx}(z) + D_{yy}(z)$;
dotted line: diffusion coefficient $D_{zz}(z)$ 
perpendicular to the interface;
thin solid line: trace $D(z)$ of the diffusion tensor in free space
(both the modified vacuum correlations 
and the curvature of the driving field are neglected).
\newline
Parameters: same as Fig.~\ref{fig:force-scalaire}. The diffusion
coefficient is plotted in units of $\hbar^2 k^2 \Gamma'_\infty$,
as a function of $kz$.}
\label{fig:trace-D}
\end{figure}
\begin{figure}
\centerline{%
\epsfxsize=0.8\columnwidth \epsfbox[150 250 450 720]{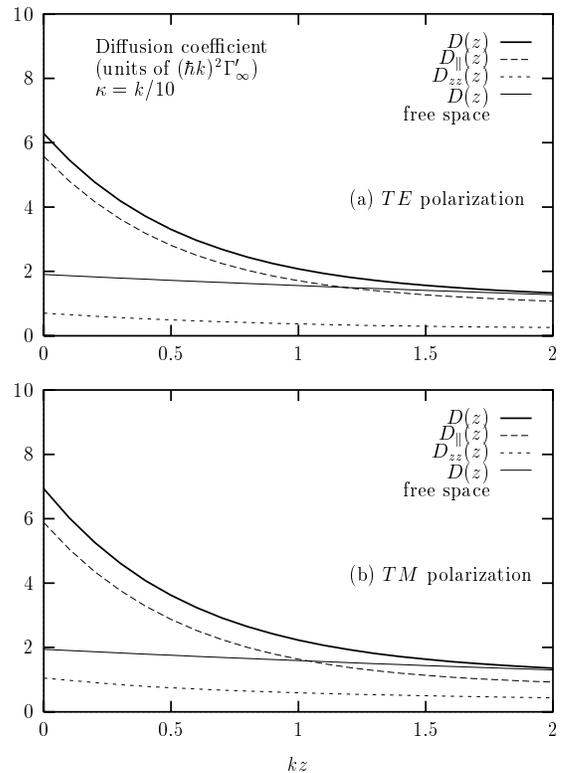}%
}
\caption[fig:trace-D2]{Same as Fig.~\ref{fig:trace-D}, but for
an evanescent driving field with larger decay length:
$\kappa = 0.1\,k$, $Q \approx 1.005\, k$.}
\label{fig:trace-D2}
\end{figure}

Quantitative results for the momentum diffusion tensor are displayed in Figs.~%
\ref{fig:trace-D}--\ref{fig:trace-D2}. In Fig.~\ref{fig:trace-D} is plotted
the trace of the diffusion tensor 
\begin{equation}
D(z)=\sum_{i}D^{i,i}(z)  \label{eq:trace-D}
\end{equation}
that permits to estimate the total width $\Delta p$ of the atomic momentum
distribution: for a spatially constant diffusion coefficient, 
\begin{equation}
\Delta p^{2}\simeq 2Dt
\end{equation}
which may be generalized to 
\begin{equation}
\Delta p^{2}\simeq 2\int \!dt\,D[\langle z(t)\rangle ]  \label{eq:delta-p2}
\end{equation}
where $\langle z(t)\rangle $ is the mean atomic trajectory in the evanescent
field's dipole potential. Eq.(\ref{eq:delta-p2}) is valid provided the
momentum diffusion is sufficiently small so that individual atomic
trajectories remain close to the mean path, an assumption that may become
questionable depending on the experimental conditions. The integral~(\ref
{eq:delta-p2}) may be estimated from the typical timescale $\tau \simeq
2M/\kappa p_{z,inc}$ for the reflection \cite{Cimmino92,Henkel94a} ($%
-p_{z,inc}$: incident atomic momentum along $z$) and the position $z_{0}$ of
the turning point of the mean path: $\Delta p^{2}\simeq 2\tau D(z_{0})$. The
diffusion coefficients plotted in Fig.~\ref{fig:trace-D} may thus be viewed
as the squared momentum width of the reflected atoms, given the interaction
time and varying the distance $z=z_{0}$ of the turning point.

From Fig.~\ref{fig:trace-D}, we observe that the atomic momentum diffusion
perpendicular to the surface (the coefficient $D^{z,z}$ represented
by the dotted curve) is, on its own, comparable to the
diffusion in free space (the thin solid curve, neglecting the curvature of
the exponential fluorescence rate and the modified vacuum correlations).
This may be compared with Fig.~\ref{fig:trace-D2} where the evanescent
driving field has a large decay length $1/\kappa $ (total internal
reflection close to the critical angle). The exponential field profile
is then essentially constant and the perpendicular momentum diffusion 
decreases. In this direction, the momentum diffusion coefficient 
$D^{z,z}$ is now determined, on the one hand,
by the spontaneous photons' recoil, $D^{z,z}_{\text{feed}}$,
and, on the other hand, by the derivatives of the fluorescence rates 
$\Gamma_{\Vert ,\perp }(z)$ that enter into $D^{z,z}_{\text{depart}}$
(\ref{ddepart}). Also, for the long-range evanescent
wave, the $TE$ and $TM$ polarizations produce a similar momentum diffusion
since they both correspond to a (nearly) linearly polarized driving field.

In conclusion, we observe that the momentum diffusion tensor is
systematically larger close to the dielectric surface compared to free space
(the thin solid lines in Figs.~\ref{fig:trace-D}, \ref{fig:trace-D2}), by a
factor of about three to four. Translating the width of the atomic momentum
distribution into a coherence length, we see that close to the surface,
spontaneous emission destroys the atomic coherence more efficiently than
expected from free-space considerations. This may have been expected
from the enhanced fluorescence rates $\Gamma _{\Vert ,\perp }(z)$
compared to $\Gamma _{\infty }$, as well as from the spatial
subwavelength structure of both the evanescent field and the vacuum field
correlations. Finally, we note that this diffusion tensor also translates
into an increased temperature limit for radiative atom traps in the vicinity
of surfaces 
\cite{Soeding95,Dalibard96b,Dowling97,Chevrollier97,Pfauetal,Mlynek97f,%
Ovchinnikov97b}.

\subsection{{$J_{g}=1/2$} atom}

\label{s:pompage}

In this section, we turn to a second application of the theory developed in
Sec.~\ref{s:GOBE} and focus on the optical pumping processes taking place in
atoms having a nontrivial Zeeman sublevel structure in their ground state.
By considering the simple case of a $J_{g}=1/2$ atom, we show that these
processes are modified both quantitatively and qualitatively in the vicinity
of the dielectric interface: since the fluorescence rates depend on both the
atom-interface distance and the atomic dipole orientation, the
optical pumping rates are modified. 
More strikingly, a net ground-state
magnetization is predicted to occur in a {\em linearly\/} polarized driving
field if the atomic recoil is taken into account.
We illustrate this effect by an explicit calculation of the radiation
pressure force.

\subsubsection{Atomic magnetization variables}

In the case of a $J_{g}=1/2$ atom, the Wigner function $W({\bf r},{\bf p})$
takes the form of a hermitian $2\times 2$ matrix describing the populations
and Zeeman coherences of the ground state. It is convenient to represent
this matrix using the vector $\bbox{\sigma}=(\sigma _{x},\sigma _{y},\sigma
_{z})$ of Pauli matrices 
\begin{equation}
W={\textstyle \frac{1}{2}}\left( w+\bbox{\sigma}\cdot {\bf J}\right) 
\label{eq:decomposer-rho}
\end{equation}
With this definition, the scalar function $w({\bf r},{\bf p})$ describes the
phase-space distribution of the total population, whereas the real vector $%
{\bf J}({\bf r},{\bf p})$ (analogous to the Bloch vector for a two-level
atom) gives the phase-space distribution of the atomic magnetization: for
unpolarized atoms, {\em e.g.}, one has ${\bf J}\equiv {\bf 0}$; for an
ensemble of atoms prepared in the sublevel $|+1/2\rangle _{z}$ with
respect to the $z$ axis, {\em e.g.}, $J_{z}({\bf r},{\bf p})=w({\bf r},{\bf p%
})$ and $J_{x,y}\equiv 0$. The evolution equations for the total population
and the magnetization vector are obtained from the G.O.B.E.\ by taking the
appropriate traces: 
\begin{mathletters}
\label{eq:extraire-w-J}
\begin{eqnarray}
w({\bf r},{\bf p}) &=&\text{Tr}\,W({\bf r},{\bf p}), \\
{\bf J}({\bf r},{\bf p}) &=&\text{Tr}\,[\bbox{\sigma}W({\bf r},{\bf p})].
\label{eq:extraire-J}
\end{eqnarray}

We next need the action of the reduced dipole operators ${\bf b}^{\pm }({\bf %
r})$ on the Wigner matrix. From the Clebsch--Gordan coefficients for the $%
J_{g}=1/2\to J_{e}=J_{g},J_{g}+1$ transitions, one finds 
\end{mathletters}
\begin{eqnarray}
{\bf b}^{-}({\bf r}) &=&\beta \,\bbox{\xi}({\bf r})+i\alpha \,\bbox{\sigma}%
\times \bbox{\xi}({\bf r}),  \label{eq:def-b-spineur} \\
&&
\begin{array}{l}
J_{e}=1/2:\qquad \beta =1/3,\quad \alpha =-1/3, \\ 
J_{e}=3/2:\qquad \beta =2/3,\quad \alpha =1/3;
\end{array}
\label{eq:def-beta-alpha}
\end{eqnarray}
while ${\bf b}^{+}({\bf r})$ is given by the hermitian conjugate. The first
term $\beta \,\bbox{\xi}({\bf r})$ in Eq.(\ref{eq:def-b-spineur}) is similar
to the reduced dipole operator for a scalar atom (it is
parallel to the driving field's polarization vector $\bbox{\xi}({\bf r})$); 
the second term $i\alpha \,%
\bbox{\sigma}\times \bbox{\xi}({\bf r})$ accounts for couplings between the
ground-state Zeeman sublevels, and therefore describes specific  multilevel
effects such as Raman couplings.\footnote{%
We refer to the paper by
Dubetsky and Berman~\cite{Berman96} for a generalization of the
decomposition~(\ref{eq:def-b-spineur}) 
of the dipole operator to the case $J_g \ge 1$. {\em Cf.}\ also 
Ref.\onlinecite{Cohen72} and references therein.
}

\subsubsection{Internal dynamics}

\label{s:spin-pompage}

As in the scalar atom situation, we start our analysis by considering the
classical optical pumping equation~(\ref{eq:OBE}) accounting for the
internal atomic dynamics. In the $J_{g}=1/2$ case, two quantities
characterize the internal atomic dynamics: the ground state light
shifts~[Eq.(\ref{Heff})] and the optical pumping rates~[Eq.(%
\ref{eq:operateur-pompage})] (we do not consider in this section the energy
level shifts induced by the interface). We discuss in some detail
the case of a circularly polarized evanescent wave and calculate the
pumping rate.

\paragraph{General.}

Let us first write down the light-shift Hamiltonian~(\ref{Heff})
(also a hermitian $2\times 2$ matrix): 
\begin{mathletters}
\label{eq:precession}
\begin{eqnarray}
H_{eff}& = &
\hbar \Delta ^{\prime }\,e^{-2\kappa z}\,{\bf b}_{0}^{+}\cdot \bbox{\xi}%
_{0} 
\nonumber\\
&=&\hbar \Delta ^{\prime }\,e^{-2\kappa z}\,(\beta |\bbox{\xi}%
_{0}|^{2}+\alpha {\bf h}\cdot \bbox{\sigma})
\label{eq:depl-lum-spin} \\
{\bf h} &=&\text{Im}\,\bbox{\xi}_{0}^{*}\times \bbox{\xi}_{0}
\label{eq:def-u}
\end{eqnarray}
where ${\bf b}_{0}^{\pm }$ is given by Eq.(\ref{eq:def-b-spineur}),
replacing $\bbox{\xi}({\bf r})$ with $\bbox{\xi}_{0}$, and where the ${\bf h}
$ vector corresponds to the helicity of the driving field. Following 
Cohen-Tannoudji and Dupont-Roc \cite{Cohen72},
one may interpret the second term of the light-shift operator~(\ref
{eq:precession}) in terms of a fictitious magnetic field parallel to
the helicity ${\bf h}$. This interpretation is supported by the equation
of motion for the atomic magnetization vector ${\bf J}$: it is obtained
using Eq.(\ref{eq:operateur-pompage}) for the
Wigner matrix $W$ that takes the following form
\end{mathletters}
\begin{eqnarray}
\left. \dot{W}_{relax}\right| _{0}& = &
\Gamma _{\infty }^{\prime }\,e^{-2\kappa z}\,C^{i,i}(z;{\bf 0})
\Big( b_{0i}^{-}Wb_{0i}^{+}
\nonumber\\
&& \quad 
-{\textstyle\frac{1}{2}}\{b_{0i}^{+}b_{0i}^{-},\,W\}\Big)   
\label{eq:pompage}
\end{eqnarray}
After some straightforward algebra with the Pauli matrices, we find from the
light-shift Hamiltonian~(\ref{eq:precession}) and Eq.(\ref
{eq:pompage}) the following equation of motion 
\begin{mathletters}
\label{eq:eq-pompage-tenseurs}
\begin{eqnarray}
&& \left. \frac{\partial {\bf J}}{\partial t}\right| _{0} 
= 2\alpha \,\Delta
^{\prime }e^{-2\kappa z}\,{\bf h}\times {\bf J}+2\alpha ^{2}\Gamma _{\infty
}^{\prime }\,e^{-2\kappa z}\,{\cal C}(z){\bf h}  \nonumber \\
&&-\,2\alpha ^{2}\Gamma _{\infty }^{\prime }\,e^{-2\kappa z}\,\Big( (\text{Tr%
}\,{\cal F})\,{\cal C}(z){\bf J}+(\text{Tr}\,{\cal C}(z))\,{\cal F}{\bf J}%
\Big)  \nonumber  \\
&&+\,2\Gamma _{\infty }^{\prime }\,e^{-2\kappa z}
\Big( \alpha ^{2}\{{\cal C}(z),\,{\cal F}\}{\bf J}
- \alpha \beta [{\cal C}(z),\,{\cal F}]{\bf J} \Big)  
\label{eq:eq-pompage}
\end{eqnarray}
We have used the following tensors 
\begin{eqnarray}
({\cal C}(z))^{i,j} &=&C^{i,j}(z;{\bf 0})=\left( 
\begin{array}{lll}
c_{\Vert }(z) & 0 & 0 \\ 
0 & c_{\Vert }(z) & 0 \\ 
0 & 0 & c_{\bot }(z)
\end{array}
\right) _{x,y,z}  \label{eq:def-tensor-C} \\
({\cal F})^{i,j} &=&\text{Re}\,\xi _{0i}^{*}\xi _{0j}
\label{eq:def-tensor-F}
\end{eqnarray}
Since the total population $w = {\rm Tr}\,W$
is conserved by optical pumping,
we have put $w=1$ in Eq.(\ref{eq:eq-pompage}). 

In the equation of motion~(\ref{eq:eq-pompage}), 
one identifies the precession of the magnetization vector ${\bf J}$ 
around the effective magnetic field vector ${\bf h}$ 
(the first term) and the feeding of the magnetization ${\bf J}$ 
through optical pumping (the second term). The third and fourth
terms describe the damping of the atomic magnetization
through absorption-spontaneous emission cycles. The effect of the
dielectric interface is encoded in the (diagonal) tensor ${\cal C}(z)$ whose
elements are proportional to the dipole damping rates $\Gamma _{\Vert }$ and 
$\Gamma _{\bot }$ [see Eq.(\ref{eq:taux-modifies})].
It is interesting to note from Eq.(\ref{eq:eq-pompage-tenseurs}) 
that optical pumping creates a magnetization 
aligned parallel to the vector ${\cal C}(z){\bf h}$
that is generally {\em not\/} parallel to the helicity ${\bf h}$ 
as a consequence of the anisotropic fluorescence rates. 
For the pumping process close to the dielectric interface, 
the one-point correlation tensor ${\cal C}(z)$
hence plays the role of an (anisotropic) `effective magnetic
susceptibility', linking the induced atomic magnetization 
to the effective magnetic field.

\paragraph{Discussion of elementary polarizations.}

According to the light-shift operator~(\ref{eq:depl-lum-spin})
and the equation of motion (\ref{eq:eq-pompage}), the
optical pumping process is characterized by the helicity vector~%
${\bf h}$ of the driving field. For the elementary polarizations
of the evanescent wave, it becomes
\end{mathletters}
\begin{mathletters}
\label{eq:depl-lum-TE/TM}
\begin{eqnarray}
\text{$TE$ polarization:} \quad
{\bf h}^{(TE)} & = & {\bf 0} 
\label{eq:depl-lum-TE} \\
\text{$TM$ polarization:} \quad
{\bf h}^{(TM)} & = & - ( 2 \kappa Q / k^2 ) {\bf e}_{y}
\label{eq:depl-lum-TM} 
\end{eqnarray}
We also mention for later use the case of a circularly 
polarized evanescent wave. As discussed at the end
of Subsec.~\ref{s:force-scalaire}, the field's polarization vector
is given by Eq.(\ref{eq:polar-sigma}) and one has
\begin{equation}
\text{circular polarization:} \quad
{\bf h}^{(\sigma)} =
( 2 Q / k ) [ {\bf e}_x - (\kappa / k) {\bf e}_y ]
\label{eq:depl-lum-sigma}
\end{equation}
In the $TE$ case, the helicity vanishes and hence no net magnetization
builds up. The light-shifts
for the two ground state Zeeman sublevels being identical (for
the same reason), one would expect
this case to yield a situation similar to 
the scalar atom of Subsec.~\ref{s:scalaire}. 
We shall see, however, that this is no longer true when the atomic recoil 
is taken into account ({\em cf.}\ Subsec.~\ref{s:recoil-induced-spin}).

In the cases of $TM$ and circular polarization, the helicity is nonzero
and optical pumping leads to a net atomic magnetization. 
In steady state, it aligns parallel to the $y$-axis in the $TM$ case, 
and parallel to an axis in the $xy$-plane for a circularly
polarized evanescent wave. In both cases, the light-shift operator
is diagonal with respect to these axis', and the anisotropic
magnetic susceptibility does not come into play. [This would be
the case for a different relative phase between the $TE$ and $TM$ 
polarizations in Eq.(\ref{eq:polar-sigma}), 
giving the helicity vector a nonzero component $h_z$.] 
\end{mathletters}

\paragraph{Example: circular polarization.}

As an example, we study in more detail the case of
a circularly polarized evanescent wave and defer the $TM$ case
to Appendix~\ref{a:pompage-TM}. 

\begin{figure}
\centerline{%
\epsfxsize=0.7\columnwidth \epsfbox[120 40 400 450]{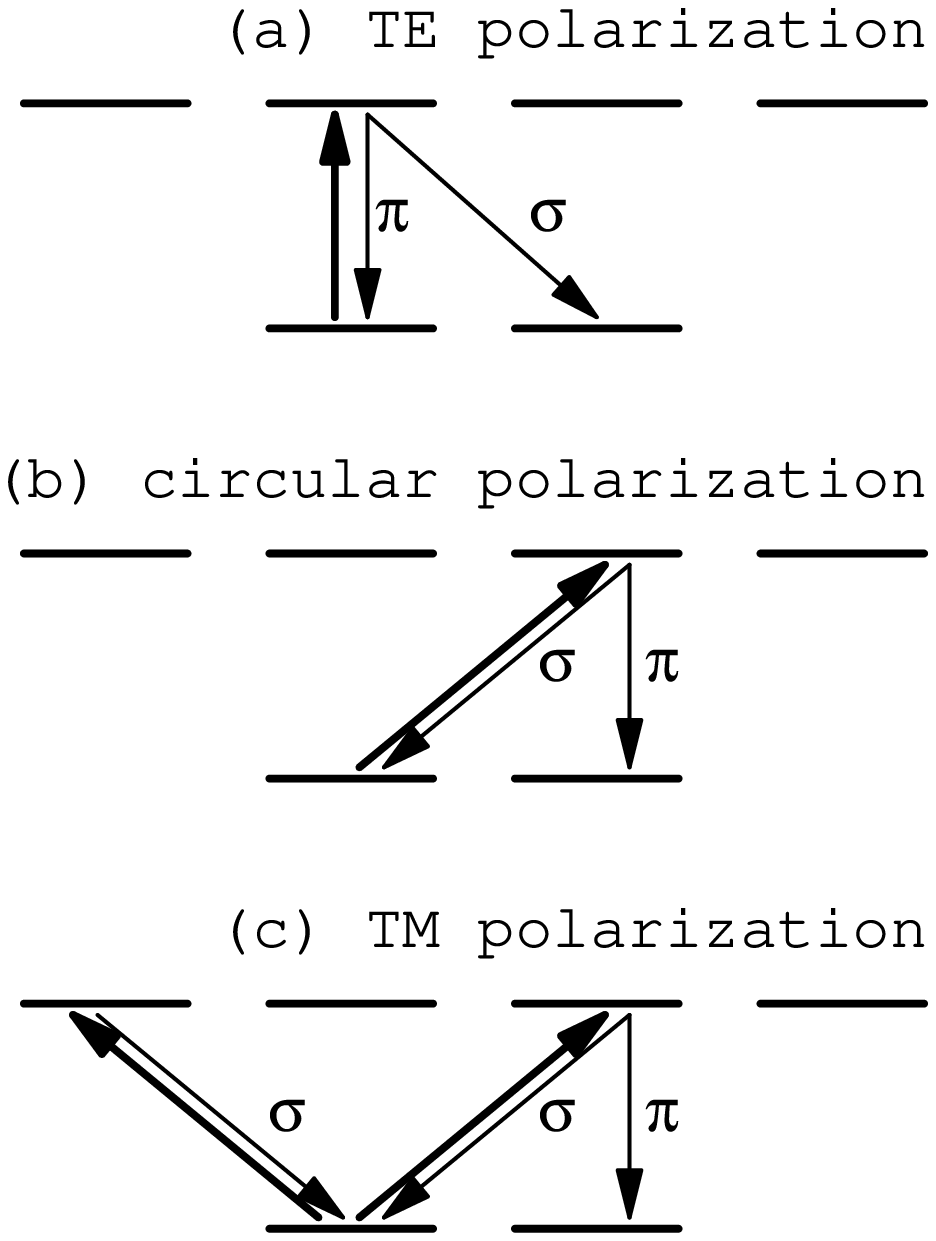}%
}
\caption[fig:cycles-fluo]{Illustration of the fluorescence cycles
for a $J = 1/2 \to J_e = 3/2$ atom 
driven by (a) a linearly polarized field ($TE$ polarization);
(b) a circularly polarized field (combination of $TE$ and
$TM$ polarization); (c) a $TM$ polarized field (elliptic polarization).
\newline
For clarity, only the fluorescence cycles starting from the sublevel
$|-1/2\rangle$ are shown. Note that these sublevels are defined
with respect to a quantization axis that, in case (b), differs
from cases (a) and (c).}
\label{fig:cycles-fluo}
\end{figure}
It is convenient to introduce a rotated coordinate system
with the ${\bf e}_{x'}$ unit vector being parallel to the 
helicity~(\ref{eq:depl-lum-sigma}). Choosing this as the
quantization axis diagonalizes in fact the light-shift 
operator~(\ref{eq:precession}). 
From the magnetization's evolution Eq.(\ref{eq:eq-pompage-tenseurs}), 
it may also be verified that
the components $J_{y'}, J_z$ decouple from $J_{x'}$. 
If we assume that the atoms are initially unpolarized, 
the pumping process only depends on $J_{x'}$ 
whose evolution is given by
\begin{mathletters}
\label{eq:pompage-circ}
\begin{equation}
\frac{ \partial J_{x'} }{ \partial t } = 
\Gamma_p( z ) [ 1 - J_{x'} ]
	\label{eq:pompage-Jx}
\end{equation}
where the `optical pumping rate' equals
\begin{equation}
\Gamma_p( z ) =  4 \alpha^2 \frac{ Q^2 }{ k^2 } 
\Gamma'_\infty e^{-2\kappa z} c_\Vert( z )
	\label{eq:taux-pompage-circ}
\end{equation}
The significance of this equation becomes evident if
we write down the evolution of the populations 
$w_\pm \equiv \frac 12 ( 1 \pm J_{x'}) $ of the Zeeman sublevels
$|\pm 1/2\rangle_{x'}$ with respect to the quantization
axis. The following rate equations are easily found
\end{mathletters}
\begin{equation}
\frac{ \partial w_+ }{ \partial t } =
\Gamma_p( z ) w_- ,
\qquad
\frac{ \partial w_- }{ \partial t } =
- \Gamma_p( z ) w_-
\label{eq:taux-circ}
\end{equation}
and we see that the pumping rate $\Gamma_p( z )$ 
governs the transition $|-1/2\rangle_{x'} \to |+1/2\rangle_{x'}$ 
between the Zeeman sublevels. 
In this transition, the atom absorbs
a $\sigma^+$ polarized photon (with respect to the $x'$ axis)
from the driving field and spontaneously emits a 
$\pi$ polarized photon ({\em cf.}\ Fig.~\ref{fig:cycles-fluo}b).
The reverse transition is impossible since the driving
field's polarization is purely $\sigma^+$.
From this elementary picture, we may understand why the optical
pumping rate (\ref{eq:taux-pompage-circ}) involves the coefficient 
$c_\Vert$ from the vacuum correlation tensor: 
the spontaneous $\pi$ photon has in fact an electric field  
parallel to the $x'$ axis, and its emission rate is
proportional to the strength of the vacuum fluctuations 
polarized along this axis, hence proportional to the
element $C^{x',x'} = c_\Vert$ of the
correlation tensor. 

The rate equations~(\ref{eq:taux-circ}) show that in steady-state, 
the atoms are completely magnetized along the helicity vector
of the driving field (the $x'$ axis). 
With respect to optical pumping in free space,
the only difference is hence the nonzero angle between this vector 
and the evanescent wave's propagation vector $Q {\bf e}_x$. 

A more detailed investigation of the pumping process may
be done in the transient regime where the interaction
time $\tau$ is smaller than the pumping time $1/\Gamma_p$.
This regime is in fact typical for atomic mirror experiments
where one seeks to avoid spontaneous emission because it
reduces the coherence of the reflection. 
An approximate solution of Eq.(\ref{eq:pompage-circ}) 
in the transient regime is (for initially unpolarized atoms)
\begin{eqnarray}
\Gamma_p \tau \ll 1 : \qquad
J_{x'} & = & w_+ - w_- \simeq \Gamma_p( z_0 ) \tau 
	\nonumber\\
& \simeq & 
4 \alpha^2 \frac{ Q^2 }{ k^2 } \Gamma'_\infty e^{-2\kappa z_0}
c_\Vert( z_0 ) \tau
\label{eq:solution-transitoire}
\end{eqnarray}
The population difference now depends on the coefficient
$c_\Vert( z_0 )$ at roughly the distance of closest approach $z_0$.
The estimate~(\ref{eq:solution-transitoire}) is actually very crude, 
since it neglects the fact that the sublevels $|\pm1/2\rangle_{x'}$ 
are subject to different light shift potentials in the circular
polarization case. This leads to a different potential (and, 
ultimately, kinetic) energy after the sublevel change---a feature that
has been studied already for both spontaneous \cite{Mlynek94b,%
Ovchinnikov95,Dalibard96c} and stimulated \cite{Ertmer93,%
Savage96,Henkel97b} transitions between sublevels. In order to
describe both the center-of-mass motion and the anisotropic
vacuum correlations, one may use the full Fokker--Planck equation 
derived in Sec.~\ref{s:GOBE}. An example of the corresponding
`recoil-induced magnetizations' is given in the next subsection.

For the sake of completeness, we have analyzed in Appendix~\ref{a:pompage-TM}
the optical pumping in a $TM$-polarized evanescent wave. 
In this case, the atoms do not become completely magnetized because the
driving field~[Eq.(\ref{eq:def-f-TE/TM})] is elliptically polarized. 
[It contains both $\sigma^+$ and $\sigma^-$ components with 
respect to the $y$-axis, {\em cf.}\ Fig.~\ref{fig:cycles-fluo}c.] 
We find a steady-state magnetization  
$J_y^{(stat)} = - 2 \kappa Q / (\kappa^2 + Q^2) > -1$,
and a pumping rate that is again proportional to the coefficient
$c_\Vert$. 

Summarizing, if atomic recoil is neglected, the internal dynamics
close to the dielectric is subject to the following modifications
as compared to free space. The optical pumping rates increase
and differ according to the polarization of the spontaneous photon 
emitted in the pumping cycle.
In a circularly polarized, evanescent driving field, a net atomic
magnetization builds up that does not align parallel 
to the field propagation vector for two reasons: 
first, the magnetization is determined 
by the field helicity that is not parallel to its wave vector, 
and second, atomic magnetization and field helicity are connected
by an anisotropic effective susceptibility because of the anisotropic 
fluorescence rates.

\subsubsection{External dynamics: recoil-induced magnetization}
\label{s:recoil-induced-spin}

We now consider the G.O.B.E. accounting for the atomic recoils during
absorption and emission of photons. Among the huge diversity of non
standard effects expected in the external dynamics of multilevel atoms in
the vicinity of a vacuum-dielectric interface, that could not be addressed
in a single paper, we focus here on radiation pressure and show
that this force may induce a net magnetization of the atomic ground state
for certain classes of the atomic velocity in a situation where the
classical optical Bloch equations would predict a zero result (linearly
polarized driving field).

\paragraph{Radiation pressure force.}

Consider the situation of a $J_{g}=1/2$ atom driven by a $TE$ polarized
evanescent wave. Using the general expressions~(\ref{eq:feed-force-general}, 
\ref{eq:depart-force-general}) for the radiation pressure force, after a
straightforward calculation using the Pauli matrices, one finds that the
departure contribution ${\bf F}_{\text{depart}}^{(sp)}$~(\ref
{eq:depart-force-general}) vanishes, whereas the feeding contribution 
${\bf F}_{\text{feed}}^{(sp)}$~(\ref{eq:feed-force-general}) 
is a sum of two terms
analogous to the ones encountered for the scalar atom [Eq.(\ref
{eq:force-scalaire})]. The first of these contributions, ${\bf F}%
_{(1)}^{(sp)}$, corresponds to the product of the fluorescence rate and the
evanescent field phase gradient. This force is hence parallel to the
evanescent wave propagation vector $Q{\bf e}_{x}$. 
We find that this term does not couple the population $w$ 
and the magnetization components ${\bf J}$. 
The second contribution to the radiation pressure operator~(\ref
{eq:feed-force-general}), ${\bf F}_{(2)}^{(sp)}$, is associated with the
gradient of the field emitted by the atomic dipole and backreflected towards
the atom by the dielectric interface, and involves the axial part of the
field correlation tensor. This contribution, that vanished for a $TE$
polarization and a $J_{g}=0\rightarrow J_{e}=1$ atom, takes a nonzero value
in the present situation and gives rise to a
magnetization--population coupling. We interpret this coupling in
terms of generalized rate equations for the Zeeman sublevel
phase-space distributions.

To be more explicit, the contribution ${\bf F}_{(1)}^{(sp)}$ 
to the radiation pressure force appears in the following way
in the Liouville equation for the magnetization component $J_i$
[{\em cf.}\ Eqs.(\ref{eq:Liouville}, \ref{eq:extraire-J})]:
\begin{equation}
\text{Tr}\,\{ \sigma_i\, 
{\bf F}^{(sp)}_{(1)}( {\bf r} ) \cdot \nabla_{\bf p} W
\}
=
{\bf F}^{(sp)}_{(1)}( {\bf r}; J_i ) \cdot \nabla_{\bf p}
J_i
  \label{eq:definir-force-1}
\end{equation}
It is characteristic for the force ${\bf F}_{(1)}^{(sp)}$  
that on the rhs only $J_i$ appears.
A similar result holds for the total population $w$. 
The following expressions are found (the $z$-dependence
of $c_\Vert, c_\perp$ has been suppressed for clarity):
\begin{mathletters}
\label{eq:force-1-spin}
\begin{eqnarray}
{\bf F}_{(1)}^{(sp)}({\bf r};w) &=& {\bf F}^{(sp,1)}
\left[ \beta ^{2}c_{\Vert } 
+ \alpha ^{2}(c_{\Vert} + c_{\perp })\right] 
\label{eq:force-1-w} \\
{\bf F}_{(1)}^{(sp)}({\bf r};J_{x}) &=& {\bf F}^{(sp,1)}
\left[ \beta ^{2}c_{\Vert }
- \alpha^{2}(c_{\Vert } - c_{\perp })\right]  \\
{\bf F}_{(1)}^{(sp)}({\bf r};J_{y}) &=& {\bf F}^{(sp,1)}
\left[ \beta ^{2}c_{\Vert }
- \alpha^{2}(c_{\Vert } + c_{\perp })\right]  \\
{\bf F}_{(1)}^{(sp)}({\bf r};J_{z}) &=& {\bf F}^{(sp,1)}
\left[ \beta ^{2}c_{\Vert }
+ \alpha^{2}(c_{\Vert } - c_{\perp })\right]  \\
{\bf F}^{(sp,1)} & = & \Gamma _{\infty }^{\prime}
\,e^{-2\kappa z}\hbar Q{\bf e}_{x} 
\end{eqnarray}
Note that the effect of the radiation pressure 
force (\ref{eq:force-1-spin}) differs between
the magnetization components. 
This feature is interpreted below using the rate 
equations~(\ref{eq:taux-force-TE})
where we show that the radiation pressure depends on whether
or not a fluorescence cycle leads to a sublevel change. 
In any case, the contribution (\ref{eq:force-1-spin})
to the radiation pressure is parallel to the propagation vector 
$Q{\bf e}_x$ of the evanescent wave.  We also observe that the effect
of the force (\ref{eq:force-1-w}) on the population $w$ is proportional 
to the total fluorescence rate $\Gamma^{\prime(TE)}( z )$:
\end{mathletters}
\begin{mathletters}
\label{eq:taux-fluo-TE-spin}
\begin{eqnarray}
{\bf F}_{(1)}^{(sp)}({\bf r};w) 
& = & \hbar Q {\bf e}_x \Gamma^{\prime(TE)}( z )
\nonumber\\
& = & \hbar Q {\bf e}_x \left( \Gamma'_\pi( z ) + 
\Gamma'_\sigma( z ) \right)
\\
\Gamma'_\pi( z ) & = &
\beta ^{2} \Gamma'_\infty e^{ -2 \kappa z }
c_{\Vert }( z ) 
\\
\Gamma'_\sigma( z ) & = &
2\alpha ^{2}
\Gamma'_\infty e^{ -2 \kappa z }
{\textstyle\frac{1}{2}}(c_{\Vert }(z)+c_{\perp }(z))  
\end{eqnarray}
The notations $\Gamma'_\pi$ and $\Gamma'_\sigma$ refer to
a quantization axis chosen along the $y$ axis
({\em cf.}\ Fig.~\ref{fig:cycles-fluo}a): with respect
to this axis, the (linearly polarized) driving field excites
a $\pi$ transition and $\Gamma'_\pi$ gives the fluorescence
rate for spontaneous photons with $\pi$ polarization
(electric field parallel to the $y$ axis). As discussed
in the example of a circular driving field,
this rate is proportional to the coefficient $c_{\Vert }$. 
We observe that the fluorescence rate $\Gamma'_\sigma$
for $\sigma$ polarized photons 
is proportional to $\frac{1}{2}(c_{\Vert }+c_{\perp })$,
these photons having an electric field in the $xz$ plane.

The contribution ${\bf F}_{(2)}^{(sp)}$ to the radiation pressure 
operator~(\ref{eq:feed-force-general}), in contrast to 
Eq.(\ref{eq:force-1-spin}), mixes the population $w$ and the
magnetization ${\bf J}$. More precisely, the Liouville equations 
for $w$, $J_x$ and $J_y$ contain the following terms 
proportional to the axial coefficient $a_1$ of the field correlation 
tensor:
\end{mathletters}
\begin{mathletters}
\label{eq:force-2-spin}
\begin{eqnarray}
&&
 \text{Tr}\, \{ {\bf F}^{(sp)}_{(2)}( {\bf r} ) \cdot \nabla_{\bf p} W \}
= 
f^{(sp,2)}\,
\Big[ \alpha \beta \frac{ \partial J_{x} }{ \partial p_{y} } 
  - \alpha ^{2} \frac{ \partial J_{y} }{ \partial p_{x} } 
\Big] 
\\
&&
\text{Tr}\, \{ \sigma_x 
( {\bf F}^{(sp)}_{(2)}( {\bf r} ) \cdot \nabla_{\bf p} W ) \} 
=
- \alpha \beta f^{(sp,2)} \frac{ \partial w }{ \partial p_{y} }
\\
&&
\text{Tr}\, \{ \sigma_y
( {\bf F}^{(sp)}_{(2)}( {\bf r} ) \cdot \nabla_{\bf p} W ) \}
=
\alpha ^{2} f^{(sp,2)} \frac{ \partial w }{ \partial p_{x} }
\label{eq:force-2-spin-Jy}\\    
&& \quad
f^{(sp,2)} 
=
2\Gamma _{\infty }^{\prime}\,e^{-2\kappa z}\hbar k a_1(z;0)
\label{eq:def-f-sp2}
\end{eqnarray}
The $J_{z}$ magnetization component is not coupled to $\{w,J_{x},J_{y}\}$
in this case.

\paragraph{Rate equations.}

In order to make the physical content of the Liouville equations~(\ref
{eq:force-1-spin}, \ref{eq:force-2-spin}) more transparent, we again 
consider initially unpolarized atoms and suppose that their momentum 
distribution is uniform in the $y$ direction (perpendicular to the 
evanescent wave's propagation vector). 
In these conditions, it is possible to neglect terms
involving the derivatives $\partial /\partial p_{y}$ in Eq.(\ref
{eq:force-2-spin}). The coupled Liouville equations then
transform into a pair
of rate equations involving only the sublevel populations 
$w_{\pm } = \frac12 (w \pm J_y)$ with respect to the $y$-axis: 
\end{mathletters}
\begin{mathletters}
\label{eq:taux-force-TE}
\begin{eqnarray}
&&
\Big( \left. \partial _{t}\right| _{0+1}+\frac{{\bf p}}{M}\cdot \nabla _{%
{\bf r}}\Big) 
w_{\pm }+\gamma _{\pm \to \mp }(z)\,w_{\pm }-\gamma _{\mp
\to \pm }(z)\,w_{\mp }
\nonumber
\\
&&
+{\bf F}_{\pm \rightarrow \pm }(z)\cdot \nabla _{{\bf p%
}}w_{\pm }+{\bf F}_{\mp \rightarrow \pm }(z)\cdot \nabla _{{\bf p}}w_{\mp } 
= 0
\label{eq:taux-wpm}
\end{eqnarray}
The different quantities in these equations are easily
found by comparison between
the Bloch~(\ref{eq:eq-pompage}) and the Liouville 
equations~(\ref{eq:force-1-spin}, \ref{eq:force-2-spin}): 
\begin{eqnarray}
\gamma _{\pm \to \mp }(z) &=& \Gamma'_{\sigma }(z)
\\
{\bf F}_{\pm \rightarrow \pm }(z) &=&
2\beta \Delta ^{\prime }\,e^{-2\kappa z}\hbar \kappa {\bf e}_{z}
+ \Gamma'_\pi( z ) \hbar Q {\bf e}_x
\label{eq:force-fluo-idem}
\\
{\bf F}_{\mp \rightarrow \pm }(z) &=&\Gamma' _{\sigma }(z)\hbar Q{\bf e}%
_{x} \mp \alpha^2 f^{(sp,2)} {\bf e}_x
\label{eq:force-fluo-diff}
\end{eqnarray}
The significance of these results is clear. The $\gamma _{\pm \to \mp }(z)$
are the transition rates for a sublevel change $|\pm 1/2\rangle _{y}\to |\mp
1/2\rangle _{y}$; both transitions take
place at the rate $\Gamma' _{\sigma }(z)$, the fluorescence rate
for $\sigma^{\pm }$ polarized spontaneous photons 
[{\em cf.}\ Eq.(\ref{eq:taux-fluo-TE-spin}) and 
Fig.~\ref{fig:cycles-fluo}a]. 

The forces ${\bf F}_{\pm \rightarrow \pm }(z)$ are the
sum of the dipole force (the first term in Eq.(\ref{eq:force-fluo-idem}))
and the radiation pressure force due to fluorescence
cycles where the atoms fall back to the same initial sublevel
(the second term). Since the
driving field is linearly polarized, this force is proportional to the
fluorescence rate $\Gamma' _{\pi}( z )$ 
for $\pi $ polarized photons ({\em cf.}\ Fig.~\ref{fig:cycles-fluo}a).
 
Finally, the forces ${\bf F}_{\mp \rightarrow \pm }$
are radiation pressure forces due to sublevel-changing fluorescence cycles $%
|\mp 1/2\rangle _{y}\to |\pm 1/2\rangle _{y}$. They differ in two respects
from the previous force. First, their mean value (averaged over the
sublevels) is proportional to the emission rate 
$\Gamma' _{\sigma }(z)$ for $\sigma_{\pm }$ polarized photons. 
Second and more striking, the forces (\ref{eq:force-fluo-diff}) are not
the same for the transitions $|+1/2\rangle _{y}\to |-1/2\rangle _{y}$ and $%
|-1/2\rangle _{y}\to |+1/2\rangle _{y}$, their 
difference being proportional
to the weight function $a_1(z;0)$ for the axial part of the field correlation
tensor [{\em cf.}\ Eq.(\ref{eq:def-f-sp2})]. 
To understand this result, we recall that for a scalar atom, the
axial part comes into play when the atom, driven
by a circularly polarized field, emits circularly polarized photons. 
More precisely, the photons must
be polarized in a plane perpendicular to the interface (the $y'z$ plane
in the example studied in the preceding paragraph). If this is the case,
the axial correlation tensor results in a force in the polarization plane
and parallel to the interface, with a sign depending on the helicity
of the spontaneous photon (see Fig.~\ref{fig:polar-sigma}). 
We encounter here a similar effect for
a $J_{g}=1/2$ atom: even in a linearly polarized driving field,
the spontaneous photon's polarization is indeed circular as soon as the atom
changes sublevel. Consider for example the transition 
$|-1/2\rangle _{y}\to |+1/2\rangle _{y}$ shown in Fig.~\ref{fig:cycles-fluo}a.
The spontaneous photon is $\sigma ^{-}$ polarized and since the quantization
axis is the $y$ axis, its electric field lies in the $xz$ plane.
The emission of this photon hence gives rise to a force
correction parallel to ${\bf e}_{x}$. On the other hand,
the reverse transition $|+1/2\rangle_{y}\to |-1/2\rangle _{y}$ 
is associated with a $\sigma ^{+}$ polarized photon and 
a force correction of opposite sign.

\begin{figure}
\centerline{%
\epsfxsize=0.7\columnwidth \epsfbox[100 0 300 140]{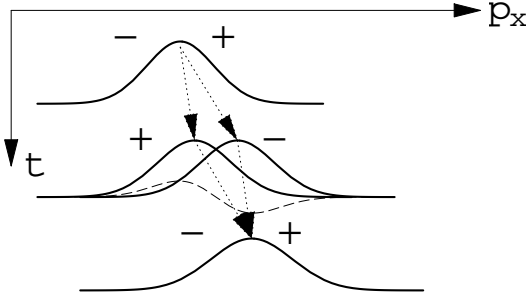}%
}
\caption[fig:oscillations]{Illustration of `recoil-induced
magnetization' in a $TE$ polarized evanescent wave. 
The momentum distributions for the sublevels
$|\pm1/2\rangle_y$ (denoted by `$+$' and `$-$') are shifted
by the radiation pressure forces ${\bf F}_{\pm}$.
Due to the difference in recoil momentum per pumping cycle,
the sublevel distributions separate and merge periodically 
at the optical pumping rate. The dashed line shows the net
magnetization $J_y(p_x)$ as a function of the momentum component
$p_x$ parallel to the evanescent wave's propagation vector.} 
\label{fig:oscillations}
\end{figure}

\paragraph{Experimental signature.}

As a consequence of the difference between the radiation pressure forces $%
{\bf F}_{\mp \rightarrow \pm }(z)$, the atomic Zeeman sublevels absorb
different momenta per optical pumping time. Their distributions hence
separate in momentum space. But since the sublevels have been exchanged
in the pumping cycle, the sublevel momentum distributions 
$w_{\pm}( p_x )$ merge
again after a second pumping cycle. The process is then repeated
periodically in time, as shown schematically in Fig.~\ref{fig:oscillations}. 
The maximum sublevel separation in momentum space is of the order of a few
photon momenta 
\end{mathletters}
\begin{equation}
\max \delta {\bf p}\simeq \frac{{\bf F}_{+\rightarrow -}^{\prime }-{\bf F}%
_{-\rightarrow +}}{\Gamma' _{\sigma }}=-4\hbar k{\bf e}_{x}\frac{a_1(z;0)}{%
c_{\Vert }(z)+c_{\perp }(z)}  \label{eq:separation-px}
\end{equation}
We note that in the transient regime, 
this phenomenon may be observed experimentally
using similar techniques as for atomic diffraction experiments at 
normal incidence \cite{Landragin97}. 
For longer interaction times, one could think of Raman
spectroscopy to detect the Zeeman sublevel imbalance as a function of the
velocity $p_{x}/M$: Eq.(\ref{eq:force-2-spin-Jy}) indeed predicts a
magnetization $w_{+}-w_{-}\equiv J_{y}$ proportional to the derivative $%
\partial w/\partial p_{x}$ of the atomic momentum distribution
(`recoil-induced magnetization'). Note that the periodic separation of the
sublevels might be difficult to observe because their momentum distributions
are also broadened by spontaneous emission (momentum diffusion).

\section{Conclusion}

We have formulated a theoretical framework to describe the motion of an atom
that is fluorescing in an environment with modified electromagnetic field
modes. We provided expressions for the radiation pressure force and the
momentum diffusion tensor in the limits of low, but semiclassical velocites
and low saturation. This general theory applies to atoms with arbitrary
Zeeman sublevel structure and environments with arbitrary electromagnetic\
field correlations. One important result is that the internal and
external (center-of-mass) dynamics of the atoms is determined by the
two-point correlation tensor of the vacuum field around the atomic position.
We then made explicit predictions for simple atoms ($J_{g}=0,\,1/2$) in the
vicinity of a flat dielectric surface. These results allow for precise
estimates of the effects of spontaneous emission when atoms are either
reflected from an evanescent wave mirror 
\cite{Ovchinnikov95,Dalibard96c,Wilkens97b}
or trapped in a two-dimensional
waveguide-like field configuration in the vicinity of a surface
\cite{Soeding95,Dalibard96b,Dowling97,Chevrollier97,Pfauetal,Mlynek97f,%
Ovchinnikov97b}.

Even for a scalar atom ($J_{g}=0$ ground state), the radiation pressure
force exhibits quantitative and qualitative changes with respect to free
space. It is increased both due to the subwavelength structure of the
evanescent driving field and the modified vacuum correlations.
In particular, the radiation pressure is no longer parallel to the
phase gradient of the driving field due to the partial reflection
at the dielectric interface. 
The optical pumping of a $J_{g}=1/2$ atom in an evanescent wave
shows similar modifications. The atomic magnetization vector
is related to the field's helicity in an anisotropic manner,
and even a linearly polarized field gives rise to an
imbalance of the atomic sublevel populations in velocity space
(`recoil-induced magnetization'). 
The sublevel-selective detection of atoms reflected 
from an evanescent wave mirror is thus a sensitive probe of
the electromagnetic field in the simple half-cavity realized by the
vacuum--dielectric interface.

The present work might be pursued in two directions: first, one could
further explore the properties of radiation pressure in evanescent waves and
consider situations beyond the simple models studied in this paper. Let us
mention some topics of particular interest: angular momenta $J_{g}\ge 1$
because such atoms are actually used in experiments; the coupling 
between Zeeman sublevel populations and coherences for more complex
field polarizations ($TM$, circular); and the momentum diffusion for
Zeeman-degenerate atoms \cite{Moelmer90}. One may also expect that
the combination of relaxation processes and the center-of-mass motion
in different light-shift potentials leads to a variety of
motion-induced magnetizations, similar to the case of conservative
couplings explored in reflection beam-splitters \cite{Ertmer93}
and diffraction gratings \cite{Savage96,Henkel97b}.
A second direction opens up if one considers different geometries: cold atoms
trapped in high-quality cavities, {\em e.g.}, are currently receiving much
interest in the fields of cavity QED and nonlinear quantum optics. The
generalized optical Bloch equations derived in Sec.~\ref{s:GOBE} 
could be used in their present form 
to study atomic motion in semiclassical driving fields 
and might be generalized to describe more
complex phenomena as, {\em e.g.}, the absorption and transmission of a probe
field or the influence of the atoms on the cavity properties.

\acknowledgments

We are indebted to A.\ Aspect, P.\ Grangier, J.-J.\ Greffet, A.\ Landragin, 
Klaus M{\o}lmer, C.\ I.\ Westbrook, and M.\ Wilkens 
for useful remarks and discussions. 
C.\ H.\ gratefully acknowledges support from Laboratoire de 
Physique des Lasers (Universit\'{e} de Paris-Nord Villetaneuse), 
Laboratoire d'Energ\'{e}tique Mol\'{e}culaire et Macroscopique, 
Combustion (Ecole Centrale Paris) and the Deutsche Forschungsgemeinschaft.

\appendix

% ---------------------------------------------------------
% app-a.tex
% ---------------------------------------------------------

\section{Derivation of the G.O.B.E. (\protect{\ref{eqnbloch}})}
\label{a:GOBE}

We outline here the derivation of the quantum-mechanical master
equation~(\ref{eqnbloch}). The only difference to the usual
treatments \cite{Agarwal75d} is the
quantization of the atomic center-of-mass motion, 
{\em i.e.}, we take care of the ordering of the atomic position 
and momentum operators ${\bf R}, {\bf P}$. 
We only present the case without
an external driving field, since this field may be easily accounted 
for by adding a commutator with the interaction Hamiltonian
$V_{AL}$ (\ref{Vav}) to the master equation of the reduced density matrix.
We also assume that the electromagnetic field is at zero temperature
(in the vacuum state), as is usual for optical frequencies.

In the interaction representation (with respect to the free
atomic Hamiltonian $H_0$ plus the free vacuum field 
Hamiltonian $H_R$), the evolution of the full 
atom + reservoir density matrix $\rho_{AR}$ is given by
\begin{equation}
\dot{ \rho }_{AR} = \frac{ 1 }{ i \hbar } 
\left[ {V}_{AR}(t), \rho_{AR} \right]
	\label{eq:von-Neumann}
\end{equation}
with the atom-field interaction given by the electric
dipole interaction
\begin{equation}
V_{AR}(t) = - {\cal D} {\bf d}(t) \cdot {\bf E}( {\bf R}, t )
	\label{eq:V-AF}
\end{equation}
Here, ${\bf D}(t) = {\cal D} {\bf d}(t)$ 
is the atomic dipole operator and 
${\bf E}( {\bf r}, t )$ the electric field operator in
the Heisenberg picture (these operators evolve in time
according to the Hamiltonian $H_0 + H_R$).
Anticipating the approximation of slowly moving atoms, we 
neglect in Eq.(\ref{eq:V-AF})
the time-dependence of the position operator
${\bf R}$ due to $H_0$ (free flight). 

To solve Eq.(\ref{eq:von-Neumann}) 
in second-order perturbation theory, 
we first re-write it as an integral equation:
\begin{equation}
\rho_{AR}( t + \Delta t ) = \rho_{AR}( t ) + 
\frac{ 1 }{ i \hbar }
\int\limits_t^{t+\Delta t} \!
d\tau \,
\left[ {V}_{AR}(\tau), \rho_{AR}( \tau ) \right]
  \label{eq:von-Neumann-integrale}
\end{equation}
This equation is iterated by inserting a similar 
expression for $\rho_{AR}( \tau )$ under the integral sign. 
Taking then the trace over the vacuum field variables, one obtains 
the master equation for the reduced atomic density matrix $\rho$. 
As usual in perturbation theory, the second-order term in ${V}_{AR}$ 
is simplified by factorizing the full density matrix according to
$\rho_{AR}( \tau ) = \rho( \tau ) \otimes \rho_{\text{vac}}$.
The resulting master equation reads in the position representation
\begin{eqnarray}
&& \langle {\bf r}_1 |
\rho( t + \Delta t ) 
| {\bf r}_2 \rangle
= 
\langle {\bf r}_1 | \rho( t ) | {\bf r}_2 \rangle 
+
\frac{ {\cal D}^2 }{ \hbar^2 }
\int\limits_t^{t + \Delta t} \! d\tau' 
\int\limits_t^{\tau'} \! d\tau
\sum\limits_{i,j = x, y, z}
	\nonumber\\
&& \Big\{
- \,
\langle 
E_i^{(+)}( {\bf r}_1, \tau' ) E_j^{(-)}( {\bf r}_1, \tau ) 
\rangle
\, d_i(\tau') d_j(\tau) 
\langle {\bf r}_1 | \rho( \tau ) | {\bf r}_2 \rangle
	\nonumber\\
&& - \,
\langle 
E_i^{(+)}( {\bf r}_2, \tau ) E_j^{(-)}( {\bf r}_2, \tau' ) 
\rangle
\, \langle {\bf r}_1 | \rho( \tau ) | {\bf r}_2 \rangle
d_i(\tau) d_j(\tau') 
	\nonumber\\
&& + \,
\langle 
E_i^{(+)}( {\bf r}_2, \tau ) E_j^{(-)}( {\bf r}_1, \tau' ) 
\rangle
\, d_j(\tau') 
\langle {\bf r}_1 | \rho( \tau ) | {\bf r}_2 \rangle
d_i(\tau) 
	\nonumber\\
&&
+ \,
\langle 
E_i^{(+)}( {\bf r}_2, \tau' ) E_j^{(-)}( {\bf r}_1, \tau ) 
\rangle
\, d_j(\tau) 
\langle {\bf r}_1 | \rho( \tau ) | {\bf r}_2 \rangle
d_i(\tau') 
\Big\} .
	\label{eq:rho-t+dt}
\end{eqnarray}
For brevity, we did not write out the vacuum state for the field 
expectation value 
$\langle 0 | E_i^{(+)}( {\bf r}_1, \tau' ) E_j^{(-)}( {\bf r}_2, \tau ) 
| 0 \rangle$ where ${\bf E}^{(\pm)}( {\bf r}, t )$ are the positive and
negative frequency components of the field operator.

We now explicit the free evolution of the atomic dipole operator for a
two-manifold system
\begin{equation}
{\bf d}(\tau) = {\bf d}^- e^{ - i \omega_A( \tau - t ) }
+
{\bf d}^+ e^{ i \omega_A( \tau - t ) },
	\label{eq:dipole-libre}
\end{equation}
where ${\bf d}^\pm$ are the dipole raising and lowering operators in
the Schr\"odinger picture at time $t$. 
We observe that the correlation time of the 
vacuum field fluctuations is much shorter
than the timescale $\Delta t$ for the evolution 
of the atomic density matrix. 
% Due to this property, the evolution of the 
% reduced density matrix becomes markovian, 
% {\it i.e.\/} the reservoir immediately 
% loses its memory of the atomic history. 
This implies that we may compute the time integrals
in Eq.(\ref{eq:rho-t+dt}) in the usual way
\cite{MandelWolf}:
replace $\tau$ by $t$ in the argument of the
density matrix and take the latter outside the integral;
change to the integration variable $\tau' - \tau$
and replace its border $\Delta t$ by infinity;
discard terms oscillating at twice the optical frequency;
identify the Fourier transform of the two-time
vacuum field correlations at the atomic frequency:
\begin{eqnarray}
&& {\cal E}^{i,j}( {\bf r}_1, {\bf r}_2 )
= 
\label{eq:fn-corr-champ}\\
&& = 
\int\limits_{-\infty}^{\infty} \! d(\tau' - \tau)
\langle 
E_i^{(+)}( {\bf r}_1, \tau' ) E_j^{(-)}( {\bf r}_2, \tau ) 
\rangle
e^{i \omega_A (\tau' - \tau) }
\nonumber
\end{eqnarray}
The remaining time integral then turns out to be proportional
to $\Delta t$, and after some term rearrangements,
one obtains the following form for the master equation 
(summation over repeated indices is understood)
\begin{eqnarray}
&& \langle {\bf r}_1 |
\frac{ \rho( t + \Delta t ) - \rho( t ) }{ \Delta t } 
| {\bf r}_2 \rangle =
\label{eq:GOBE}\\
&&  = 
- \,
\frac{ {\cal D}^2 }{ 2\hbar^2 }
\langle {\bf r}_1 |
\Big\{
{\cal E}^{i,j}( {\bf R}, {\bf R} )
d_i^+ d_j^- , \, 
 \rho( t ) 
\Big\}
| {\bf r}_2 \rangle
	\nonumber\\
&& \quad +\,
\frac{ {\cal D}^2 }{ \hbar^2 }
{\cal E}^{i,j}( {\bf r}_2, {\bf r}_1 )
d_j^-
\langle {\bf r}_1 | \rho( t ) | {\bf r}_2 \rangle
d_i^+ 
	\nonumber\\
&& \quad +\,
\frac{ 1 }{ i \hbar }
\langle {\bf r}_1 |
\left[
( H_A( {\bf R} ) - H_0 ), \, 
\rho( t )
\right]
| {\bf r}_2 \rangle
	\nonumber
\end{eqnarray}
We may identify the lhs of this equation
with the time derivative $\langle {\bf r}_1 |
\dot{ \rho } | {\bf r}_2 \rangle$ since the reduced 
density matrix evolves slowly on the time scale of the
vacuum fluctuations. 

We finally recall that the correlation 
function~(\ref{eq:fn-corr-champ})
is identical, up to a normalization,
to the field correlation tensor defined in Eq.(\ref{eq:def-fn-corr}):
$({\cal D}^2 / \hbar^2) {\cal E}^{i,j} = \Gamma_\infty C^{i,j}$.
The first two lines of Eq.(\ref{eq:GOBE}) 
are then readily identified with the relaxation part 
$\dot{ \rho }_{relax}$ of the G.O.B.E.\ (\ref{eqnbloch}).
Furthermore, the last line of Eq.(\ref{eq:GOBE})
contains the level shifts (Lamb-shifts)
due to the coupling to the reservoir. These shifts 
contain both the renormalization of the
atomic Hamiltonian, $H_{A,\infty} - H_0$, and
its the interface-dependent part $\Delta H_A( {\bf R} )$ 
appearing in Eq.(\ref{modifeqbloch}). For simplicity,
we do not write down explicit expressions and
refer to Ref.\cite{Courtois96} for a discussion of
the atomic level shifts in the vicinity of a vacuum--dielectric
interface.

% end of app-a.tex

% ---------------------------------------------------------
% app-b.tex
% ---------------------------------------------------------

\section{Adiabatic elimination of the optical coherences and
the excited state population}
\label{a:elimination}

In this appendix, we analyze in detail the validity conditions
for the adiabatic elimination of the optical coherences and
the excited state density matrix.

\subsection{Optical coherences}

The atoms are driven by a laser field that we describe by a 
monochromatic classical field
\begin{equation}
{\cal {\vec E}}_L( {\bf r}, t ) =
{\cal {\vec E}}_L( {\bf r} ) e^{ - i \omega_L t } +
\text{c.c.}
  \label{eq:champ-laser}
\end{equation}
The interaction Hamiltonian~(\ref{Vav}) becomes, in the rotating
wave approximation,
\begin{eqnarray}
V_{AL} & = & 
- {\cal D} \left(
{\bf d}^+. {\cal {\vec E}}_L( {\bf r} ) 
e^{ - i \omega_L t } +
{\bf d}^-. {\cal {\vec E}}_L^*( {\bf r} ) 
e^{ i \omega_L t }
\right)
\nonumber\\
& = &
- {\cal D} {\cal E}_0 \left(
{\bf d}^+. \bbox{\xi}( {\bf r} ) 
e^{ - i \omega_L t } +
{\bf d}^-. \bbox{\xi}^*( {\bf r} ) 
e^{ i \omega_L t }
\right)
\end{eqnarray}
where the dimensionless vector $\bbox{\xi}( {\bf r} )$ for the
field profile [Eq.(\ref{def-xi})] has been used.
The time-dependence of the interaction Hamiltonian is
removed by passing into the ``rotating frame'', {\em i.e.},
we write the optical coherence in the form
\begin{equation}
\rho_{eg} = e^{ - i \omega_L t } \tilde{\rho}_{eg}
  \label{eq:repere-tournant}
\end{equation}
The equation of motion for $\tilde{\rho}_{eg}$ is now
readily obtained from the Bloch 
equations~(\ref{Liouvillian}, \ref{equaBlochgen},
\ref{modifeqbloch}, \ref{eqnbloch})
and reads
\begin{eqnarray}
&& \bigg( 
\frac{ d }{ dt } 
- i \Delta
+ \frac{ \Gamma_\infty }{ 2 } {\cal G}_e( {\bf R} ) 
\bigg)
\tilde{ \rho }_{eg} =
	\label{eq:coh-opt}\\
&& = \frac{ 1 }{ i \hbar} 
\bigg[
P_e \Delta H_A( {\bf R} ) P_e  \tilde{ \rho }_{eg}
- \tilde{ \rho }_{eg} P_g \Delta H_A( {\bf R} ) P_g
\bigg] +
\nonumber\\ 
&& +
\frac{ 1 }{ i\hbar } 
\bigg[ \frac{ {\bf P}^2 }{ 2 M} , \,
\tilde{ \rho }_{eg} 
\bigg]
+ \frac{ i {\cal D} {\cal E}_0 }{ \hbar } 
\Big(
[ {\bf d}^+ . \bbox{\xi}( {\bf R} ) ] \sigma
- \rho_{ee} [ {\bf d}^+ . \bbox{\xi}( {\bf R} ) ]
\Big)
	\nonumber
\end{eqnarray}
The ordering of the terms takes into account 
that ${\bf R}$ is the atomic position operator.
We have also introduced the abbreviation
\begin{equation}
{\cal G}_e( {\bf R} ) \equiv
C^{i,j}( {\bf R}, {\bf R} ) d^+_i d^-_j
  \label{eq:def-Pe}
\end{equation}
Eq.(\ref{eq:coh-opt}) may be approximately solved if the detuning
$\Delta$ is outweighing all the other frequencies.
We therefore assume that the atoms are driven off-resonantly 
and at low saturation, $|\Delta| \gg \Gamma_\infty, 
{\cal D} {\cal E}_0 / \hbar, |\Delta H_A / \hbar|$, as in 
condition~(\ref{eq:cond-eliminer-coherences}).
The frequency associated with the kinetic energy operator
may be estimated in the Wigner representation
[{\em cf.}\ Eq.(\ref{eq:Liouville})].
If we assume that the atomic position distribution
varies at most on the scale of the optical wavelength,
this term is overestimated by the Doppler shift $k p / M$. 
In this way, we find the third condition appearing 
in~(\ref{eq:cond-eliminer-coherences}),
$|\Delta| \gg k p / M$.

Given these conditions,
the adiabatic solution to Eq.(\ref{eq:coh-opt}),
correct to first order in $\Gamma_\infty / \Delta$, 
reads
\begin{eqnarray}
\tilde{ \rho }_{eg} & \simeq & 
-
\frac{ {\cal D} {\cal E}_0 }{ \hbar \Delta }
\Big( 1 - i \frac{ \Gamma_\infty }{ 2 \Delta }
{\cal G}_e( {\bf R} ) \Big) 
\times \label{eq:solution-coh-opt}\\ 
&& \times \Big(
[ {\bf d}^+ . \bbox{\xi}( {\bf R} ) ] \sigma
- 
\rho_{ee} [ {\bf d}^+ . \bbox{\xi}( {\bf R} ) ]
\Big)
\nonumber
\end{eqnarray}
Note that the optical coherence $\tilde{ \rho }_{ge}$ is equal to the hermitian
conjugate of~(\ref{eq:solution-coh-opt}). We shall see in the
next paragraph that the excited state density matrix
$\rho_{ee}$ is much smaller than the ground state
density matrix $\sigma$ [Eq.(\ref{eq:solution-excite})]. 
We may therefore neglect the former in 
Eq.(\ref{eq:solution-coh-opt}). Using the
definition~(\ref{eq:def-operateur-b}) of the reduced dipole
operator ${\bf b}^-( {\bf R} )$ and recalling that
the projection operator $\sum_i d^+_i d^-_i = P_e$ acts as
the identity onto the excited state manifold, one sees that
Eq.(\ref{eq:solution-coh-opt}) yields 
the expression~(\ref{eq:coherences}) for the optical 
coherences.

\subsection{Excited state}

The generalized optical Bloch equation for the excited state 
density matrix $\rho_{ee}$ reads
\begin{eqnarray}
&& \dot{ \rho }_{ee} 
+ \frac{ \Gamma_\infty }{ 2 } 
\bigg\{
{\cal G}_e( {\bf R} ), \,
\rho_{ee}
\bigg\} =
\nonumber\\
&& = 
\frac{ 1 }{ i\hbar } 
\bigg[ \frac{ {\bf P}^2 }{ 2 M} 
+ P_e \Delta H_A( {\bf R} ) P_e , \,
\rho_{ee} 
\bigg] + 
\nonumber\\
&& 
+ \frac{ i {\cal D} {\cal E}_0 }{ \hbar } 
\Big(
[ {\bf d}^+ . \bbox{\xi}( {\bf R} ) ] \tilde{ \rho }_{ge}
- \tilde{ \rho }_{eg} [ {\bf d}^- . \bbox{\xi}^*( {\bf R} ) ]
\Big)
	\label{eq:Bloch-excite}
\end{eqnarray}
Inserting the adiabatic solution~(\ref{eq:solution-coh-opt})
for the optical coherences, one obtains
\begin{eqnarray}
&&
\dot{ \rho }_{ee} 
+ \frac{ \Gamma_\infty }{ 2 } 
\bigg\{
{\cal G}_e( {\bf R} ), \,
\rho_{ee}
\bigg\}
 =
\frac{ 1 }{ i\hbar }
\bigg[
\frac{ {\bf P}^2 }{ 2 M }, \,
\rho_{ee} \bigg]
+ \nonumber\\
&& \quad
+ \frac{ 1 }{ i\hbar }
\bigg[
P_e \Delta H_A( {\bf R} ) P_e
- \hbar\Delta \frac{ s_0 }{ 2 } 
[{\bf d}^+ . \bbox{\xi}( {\bf R} ) ] 
[{\bf d}^- . \bbox{\xi}^*( {\bf R} )], \,
\rho_{ee} \bigg]
\nonumber\\
&& \quad 
+ \frac{ \Gamma_\infty }{ 2 } \frac{ s_0 }{ 2 }
\bigg\{
{\cal G}_e( {\bf R} ), \,
[{\bf d}^+ . \bbox{\xi}( {\bf R} ) ] 
\sigma
[{\bf d}^- . \bbox{\xi}^*( {\bf R} )]
\bigg\}
	\nonumber\\
&& \quad
- \frac{ \Gamma_\infty }{ 2 }
\frac{ s_0 }{ 2 }
\Big(
[{\bf d}^+ . \bbox{\xi}( {\bf R} ) ] 
[{\bf d}^- . \bbox{\xi}^*( {\bf R} )]
\rho_{ee} {\cal G}_e( {\bf R} ) +
\nonumber\\
&& \qquad 
+ {\cal G}_e( {\bf R} ) \rho_{ee} 
[{\bf d}^+ . \bbox{\xi}( {\bf R} ) ] 
[{\bf d}^- . \bbox{\xi}^*( {\bf R} )]
\Big)
	\label{eq:excite}
\end{eqnarray}
We have used the saturation parameter $s_0$ defined in 
Eq.(\ref{eq:def-saturation}).

To solve this equation approximately, we observe that
the last term on the rhs is small compared 
to the term involving $\Gamma_\infty$ on the lhs,
because of the low saturation limit $s_0 \ll 1$
[condition~(\ref{eq:cond-eliminer-coherences})]. 
We also want to neglect the first two lines on the rhs
(the kinetic and potential energy operators in the commutator). 
We noted above that the kinetic energy corresponds to
a rate smaller than roughly the Doppler shift.
It is hence negligible if condition~(\ref{eq:atomes-lents}),
$\Gamma_\infty \gg kp / M$, holds. 
We note that this condition may be re-written as 
\begin{equation}
\frac{ 1 }{ \Gamma_\infty } 
\frac{ p }{ M } 
\ll
\lambdabar
\end{equation}
{\em i.e.}, the atoms move much less than a wavelength during
the lifetime of the excited state.
To estimate the potential energy term, we again use the
Wigner representation and find [{\em cf.}\ Eq.(\ref{eq:Liouville})]
that it is of the order of the force $F_e$ 
acting on the excited state divided by the width $\Delta p$ 
of the atomic momentum distribution.
We therefore need to suppose 
$\Gamma_\infty \gg  F_e / \Delta p $ 
[{\em cf.}\ Eq.(\ref{eq:atomes-lents})] or
\begin{equation}
\frac{ F_e }{ \Gamma_\infty } \ll
\Delta p
  \label{eqa:force-petite}
\end{equation}
The momentum the atoms gain during the excited state's lifetime
is hence negligible compared to the width of the momentum distribution.

Given these conditions, we are left with the equation
\begin{equation}
\bigg\{
{\cal G}_e( {\bf R} ), \,
\rho_{ee}
- 
\frac{ s_0 }{ 2 }
[{\bf d}^+ . \bbox{\xi}( {\bf R} ) ] 
\sigma
[{\bf d}^- . \bbox{\xi}^*( {\bf R} )]
\bigg\} \simeq 0
  \label{eq:anticommutateur}
\end{equation}
Using the fact that the spontaneous emission rates 
are positive for any polarization of the spontaneous
photon, it is easy to prove that the solution 
to~(\ref{eq:anticommutateur}) is given by
\begin{equation}
\rho_{ee} \simeq
\frac{ s_0 }{ 2 }
[{\bf d}^+ . \bbox{\xi}( {\bf R} ) ] 
\sigma
[{\bf d}^- . \bbox{\xi}^*( {\bf R} )]
	\label{eq:solution-excite}
\end{equation}
This expression shows that the excited state density matrix is
much smaller than that of the ground state, by a factor of the
order of the saturation parameter $s_0$. In the position
representation, Eq.(\ref{eq:solution-excite}) yields the
result~(\ref{eq:rho-excite}).

\subsection{Ground state}

The Bloch equation for the ground state density matrix
$\sigma$ reads in the position representation
\begin{eqnarray}
&& \langle {\bf r}_1 | 
\dot{ \sigma } 
| {\bf r}_2 \rangle 
= 
\frac{ 1 }{ i\hbar } 
\langle {\bf r}_1 | 
\bigg[ \frac{ {\bf P}^2 }{ 2 M} +
P_g \Delta H_A( {\bf R} ) P_g , \,
\sigma
\bigg]
| {\bf r}_2 \rangle 
\nonumber
\\
&&
+ \frac{ i {\cal D} {\cal E}_0 }{ \hbar } 
\Big(
[ {\bf d}^- . \bbox{\xi}^*( {\bf r}_1 ) ] 
\langle {\bf r}_1 | 
\tilde{ \rho }_{eg}
| {\bf r}_2 \rangle 
- 
\langle {\bf r}_1 | 
\tilde{ \rho }_{ge} 
| {\bf r}_2 \rangle 
[ {\bf d}^+ . \bbox{\xi}( {\bf r}_2 ) ]
\Big)
\nonumber
\\
&&
+ \Gamma_\infty
C^{i,j}( {\bf r}_2, {\bf r}_1 )
d^-_j
\langle {\bf r}_1 | 
\rho_{ee}
| {\bf r}_2 \rangle 
	\label{eq:fondamental}
\end{eqnarray}
We now insert the adiabatic expressions~(\ref{eq:solution-coh-opt},
\ref{eq:solution-excite}) for the optical coherences 
$\tilde{ \rho }_{eg}$ and the excited state density matrix
$\rho_{ee}$. The last line of Eq.(\ref{eq:fondamental}) 
readily yields the last line of the optical pumping 
equation~(\ref{eqnpompage}). In the second line, involving
the optical coherences, we neglect terms of order $s_0^2$
and obtain the following two contributions
\begin{equation}
\frac{ 1 }{ i \hbar }
\langle {\bf r}_1 |
\Big[
H_{ls}( {\bf R} ), \,
\sigma
\Big]
| {\bf r}_2 \rangle
-
\frac{ \Gamma_\infty' }{ 2 }
\langle {\bf r}_1 |
\Big\{
{\cal G}( {\bf R} ), \,
\sigma
\Big\}
| {\bf r}_2 \rangle
  \label{eq:deux-termes}
\end{equation}
The commutator is due to the real part
of the optical coherences~(\ref{eq:solution-coh-opt})
(in phase with the driving field)
and is characterized by the light-shift Hamiltonian
\begin{eqnarray}
H_{ls}( {\bf R} ) & = & 
\hbar \Delta \frac{ s_0 }{ 2 }
[ {\bf d}^- . \bbox{\xi}^*( {\bf R} ) ] 
[ {\bf d}^+ . \bbox{\xi}( {\bf R} ) ]
\nonumber\\ 
& = &
\hbar \Delta'
[ {\bf d}^- . \bbox{\xi}^*( {\bf R} ) ] 
[ {\bf d}^+ . \bbox{\xi}( {\bf R} ) ].
\end{eqnarray}
This Hamiltonian adds to the level shift
$P_g \Delta H_A( {\bf R} ) P_g$ in Eq.(\ref{eq:fondamental})
to give the effective ground-state Hamiltonian
$H_{eff}( {\bf R} )$ (\ref{Heff}).
The anticommutator in Eq.(\ref{eq:deux-termes})
is due to the imaginary part of the optical coherences
$\tilde{ \rho }_{eg}$
(phase lag of order $\Gamma_\infty / \Delta$).
It involves the ground-state operator ${\cal G}( {\bf R} )$
defined in Eq.(\ref{opg}):
\begin{eqnarray}
\frac{ \Gamma'_\infty }{ 2 }
{\cal G}( {\bf R} )
& = & 
\frac{ \Gamma'_\infty }{ 2 }
C^{i,j}( {\bf R}, {\bf R})
b^-_i( {\bf R} ) b^-_j( {\bf R} )
\nonumber\\
& = &
\frac{ \Gamma_\infty }{ 2 }
\frac{ s_0 }{ 2 }
[ {\bf d}^- . \bbox{\xi}^*( {\bf R} ) ] 
{\cal G}_e( {\bf R} )
[ {\bf d}^+ . \bbox{\xi}( {\bf R} ) ]
\end{eqnarray}
We have thus obtained the optical pumping 
equation~(\ref{eqnpompage}).

% ------- end of app-b.tex ----------------

% ---------------------------------------------------------
% app-c.tex
% ---------------------------------------------------------

\section{Field correlations for the vacuum--dielectric interface}
\label{a:fn-corr}

In this appendix, we outline the calculation of the electromagnetic
field correlation tensor for the vacuum--dielectric interface,
following Carnaglia and Mandel \cite{Mandel71}.
We assume that the dielectric fills the half-space $z < 0$ and
is characterized by the (real) refractive index $n_0$.

\subsection{Field modes}

Carnaglia and Mandel distinguish
two types of electromagnetic field modes for this geometry:

(a)
modes incident from inside the dielectric and being partially
or totally reflected at the dielectric--vacuum interface.
In the vacuum half-space, these modes are either propagating
or evanescent, depending on the internal angle of incidence. 
The wavevector of the incident wave in the dielectric
is denoted ${\bf k}_{0\uparrow}$, and the vacuum wavevector
of the transmitted wave ${\bf k}_{\uparrow}$. Obviously,
one has $| {\bf k}_{0\uparrow} |= n_0 (\omega / c) \equiv n_0 k$ and
$| {\bf k}_\uparrow | = k$.   
The wavevectors are decomposed according to ${\bf k}_\uparrow =
( {\bf k}_\Vert, k_z )$ where ${\bf k}_\Vert$
denotes the components parallel to the interface plane
(the $xy$-plane) and $k_z$ the perpendicular component.
The parallel components of ${\bf k}_0$ and ${\bf k}$ 
coincide: ${\bf k}_{0\Vert} = {\bf k}_{\Vert}$.
The perpendicular components are such that in the dielectric,
$k_{0z} > 0$; in vacuum, $k_{z}$ is chosen such that
$k_{z} > 0$ for the propagating modes 
and Im $k_{z} > 0$ for the evanescent waves.
For later convenience,
we introduce the abbreviation $ u \equiv |{\bf k}_\Vert| / k $.
Modes propagating (evanescent) in vacuum then correspond to 
$0 \le u \le 1$ ($1 < u < n_0$), respectively.

We write $ {\bf f}_{\uparrow}({\bf k}_{0\uparrow}, \mu; {\bf r}) $ 
($\mu = TE,\, TM$) for the corresponding mode function 
that is normalized to unit (incident) amplitude 
in the dielectric.
In the vacuum half-space, these modes
have an amplitude equal to the Fresnel transmission coefficient
denoted by $t(u, \mu)$. Explicitly, one has ($z > 0$): 
\begin{eqnarray}
{\bf f}_{\uparrow}({\bf k}_0, \mu; {\bf r}) & = &
{\bf e}_{\uparrow}(u, \varphi, \mu) t(u, \mu) 
\exp{( i {\bf k}_\uparrow\cdot{\bf r} ) },
\label{eq:up-mode}
\\
\text{with} \quad
{\bf k} & =  &
k ( u \cos\varphi, u \sin\varphi, v ),
\\
&& \quad ( 0 \le u \le n_0 ),
\nonumber\\
v & = & \sqrt{ 1 - u^2 },
\\ 
{\bf e}_{\uparrow}(u, \varphi, TE) & = &
( -\sin\varphi, \cos\varphi, 0 ), 
\\
{\bf e}_{\uparrow}(u, \varphi, TM) & = &
( v \cos\varphi, v \sin\varphi, -u ),
\label{eqa:polar-TM}
\\
t(u, TE) & = &
\frac{ 2 \sqrt{ n_0^2 - u^2 } }{ 
v + \sqrt{ n_0^2 - u^2 } }, 
\\
t(u, TM) & = &
\frac{ 2 n_0^2 \sqrt{ n_0^2 - u^2 } }{ 
n_0^2 v + \sqrt{ n_0^2 - u^2 } } .
\end{eqnarray}
Note that the $TM$ polarization vector (\ref{eqa:polar-TM}) is
complex for evanescent modes ($u > 1$).

(b)
modes propagating downwards from the upper half-space
into the dielectric with wavevectors
${\bf k}_\downarrow = ( {\bf k}_\Vert, - k_z )$ in vacuum
and ${\bf k}_{0\downarrow} = ( {\bf k}_\Vert, - k_{0z} )$
in the dielectric.  In the vacuum half-space,
these modes contain a part
reflected from the interface with wavevector 
${\bf k}^{(r)}_\downarrow = {\bf k}_\uparrow$.
We note ${\bf f}_{\downarrow}({\bf k}_\downarrow, \mu; {\bf r})$ 
the corresponding mode function 
(normalized to unit incident amplitude in vacuum)
and $r(u, \mu)$ the Fresnel amplitude reflection coefficient
($z > 0$):
\begin{eqnarray}
{\bf f}_{\downarrow}({\bf k}, \mu; {\bf r}) & = &
\exp{( i {\bf k}_\Vert\cdot{\bf r}_\Vert )} (
{\bf e}_{\downarrow}(u, \varphi, \mu) \, e^{- i k_z z}
+ 
\nonumber\\
&& + 
{\bf e}^{(r)}_{\downarrow}(u, \varphi, \mu) \, r(u, \mu)  e^{i k_z z}
),
\label{eq:down-mode}
\\
\text{with} \quad
{\bf k}_\downarrow & =  &
k \left( u \cos\varphi, u \sin\varphi, - v  \right),
\\
&& \quad
( 0 \le u \le 1 ),
\nonumber\\
{\bf e}_{\downarrow}(u, \varphi, TE) & = &
( -\sin\varphi, \cos\varphi, 0 )
\\
& = & {\bf e}^{(r)}_{\downarrow}(u, \varphi, TE) 
= {\bf e}_{\uparrow}(u, \varphi, TE) 
\nonumber\\
{\bf e}_{\downarrow}(u, \varphi, TM) & = &
( v \cos\varphi, v \sin\varphi, u )
\\
{\bf e}^{(r)}_{\downarrow}(u, \varphi, TM) & = &
( -v \cos\varphi, -v \sin\varphi, u )
\\
& = & - {\bf e}_{\uparrow}(u, \varphi, TM) 
\nonumber\\
r(u, TE) & = &
\frac{ v - \sqrt{ n_0^2 - u^2 } }{ 
v + \sqrt{ n_0^2 - u^2 } }
\\
r(u, TM) & = &
\frac{ n_0^2 v - \sqrt{ n_0^2 - u^2 } }{ 
n_0^2 v + \sqrt{ n_0^2 - u^2 } }
\end{eqnarray}

For a detailed discussion of the orthonormalization of these 
field modes, {\em cf.}\ Ref.~\onlinecite{Mandel71}.

\subsection{Vacuum field correlation function}

Upon quantization, the electric field operator in the vacuum half-space
may be written as a sum over the two types of modes introduced
above, the mode functions being multiplied by the usual creation
and annihilation operators \cite{Mandel71,Glauber91}.
Using the bosonic commutation rules, Carnaglia and Mandel
obtain the following result for the vacuum correlation tensor 
($z_1, z_2 > 0$):
\begin{eqnarray}
&& 
\langle 0 |
E_i^{(+)}( {\bf r}_1, \tau )\, E_j^{(-)}( {\bf r}_2, 0 )
| 0 \rangle =
\label{eq:corr-general}\\
&&
= 
\int_{>}\! \frac{ d^3 k_{0\uparrow} }{ (2\pi)^3 }
\frac{ \hbar \omega }{ 2 n_0^2 \varepsilon_0 }
\sum_\mu
f_{\uparrow i}({\bf k}_{0\uparrow}, \mu; {\bf r}_1)
f_{\uparrow j}^*({\bf k}_{0\uparrow}, \mu; {\bf r}_2)
\, e^{ - i \omega \tau }
	\nonumber\\
& & + \, 
\int_{<}\! \frac{ d^3 k_\downarrow }{ (2\pi)^3 }
\frac{ \hbar \omega }{ 2 \varepsilon_0 }
\sum_\mu
f_{\downarrow i}({\bf k}_\downarrow, \mu; {\bf r}_1)
f_{\downarrow j}^*({\bf k}_\downarrow, \mu; {\bf r}_2)
\, e^{ - i \omega \tau }
	\nonumber
\end{eqnarray}
The signs ``$>$, $<$'' on the integral signs are to remind that the
wavevectors ${\bf k}_{0\uparrow}$ and ${\bf k}_{\downarrow}$
only run through a half-space.

From expression~(\ref{eq:corr-general}), we may readily 
read off the Fourier transform
${\cal E}^{i,j}( {\bf r}_1, {\bf r}_2 )$ of the correlation
tensor defined in Eq.(\ref{eq:fn-corr-champ}), by using as
integration variables the frequency $\omega$ and the 
polar coordinates $u, \varphi$ of the in-plane vector 
${\bf k}_\Vert / k$. After some algebra, one obtains the
following representation of the normalized correlation 
tensor (\ref{eq:def-fn-corr}): 
\begin{eqnarray}
&& C^{i,j}( {\bf r}_1, {\bf r}_2 ) =
\label{eq:corr-u-du}\\
&& = 
\frac{ 3 }{ 8\pi }
\int\limits_{ 0 }^{ n_0 } \! \frac{ du \, u }{ \sqrt{ n_0^2 - u^2 } } 
\int\limits_{ 0 }^{ 2\pi } \! d\varphi 
\sum_\mu
f_{\uparrow i}({\bf k}_{0\uparrow}, \mu; {\bf r}_1)
f_{\uparrow j}^*({\bf k}_{0\uparrow}, \mu; {\bf r}_2)
\, +
	\nonumber\\
& & + \, 
\frac{ 3 }{ 8\pi }
\int\limits_{ 0 }^{ 1 } \! \frac{ du \, u }{ \sqrt{ 1 - u^2 } } 
\int\limits_{ 0 }^{ 2\pi } \! d\varphi
\sum_\mu
f_{\downarrow i}({\bf k}_\downarrow, \mu; {\bf r}_1)
f_{\downarrow j}^*({\bf k}_\downarrow, \mu; {\bf r}_2)
	\nonumber
\end{eqnarray}
[From here on, the wavevectors ${\bf k}_{\uparrow,\downarrow}$
have magnitude $| {\bf k}_{\uparrow,\downarrow} | = \omega_0 / c = k$
where $\omega_0$ is the atomic transition frequency.]
In the following, we show that the correlation function
(\ref{eq:corr-u-du})
may be written as a sum of two parts, $C = C_{\infty} + C_{int}$,
one corresponding to the free-space correlation function,
and the other one representing the influence of the interface.

\subsubsection{Free-space part}

The free-space correlation function is obtained from
those upward propagating modes 
that are homogeneous plane waves above the dielectric,
on the one hand,
and from either the incident or the reflected parts
of the downward propagating modes, on the other.
These contributions may be combined using
the following property of the Fresnel coefficients
\begin{equation}
u < 1: \qquad
\sqrt{ \frac{ 1 - u^2  }{ n_0^2 - u^2 } } t^2(u, \mu)  
+ r^2(u, \mu)  =  1
	\label{eq:reciprocity}
\end{equation}
that follows from the relation $r = -r'$ 
where $r'$ is the reflection coefficient for upward propagating modes 
(reciprocity), and energy conservation. One finally gets the result
\begin{eqnarray}
C^{i,j}_{\infty}( {\bf r}_1, {\bf r}_2 ) & = &
\frac{ 3 }{ 8 \pi }
\int\limits_{ 0 }^{ 1 } \! du
\int\limits_{ 0 }^{ 2\pi } \! d\varphi
\frac{ u }{ \sqrt{ 1 - u^2 } }
\sum_{ \mu }
\nonumber\\
&& \Big(
e_{\uparrow i}( u, \varphi, \mu ) 
e^*_{\uparrow j}( u, \varphi, \mu )
e^{ i k_z (z_1 - z_2) } \, +
\nonumber 
\\
&& \quad + \,
e_{\downarrow i}( u, \varphi, \mu ) 
e^*_{\downarrow j}( u, \varphi, \mu )
e^{ - i k_z (z_1 - z_2) }
\Big) 
\times \nonumber \\
&& \times
\exp[ i {\bf k}_\Vert\cdot( {\bf r}_{\Vert,1} - {\bf r}_{\Vert,2} ) ]
	\label{eq:corr-vide-angle-solide}
\end{eqnarray}
If the integral is written in terms of the unit vector
${\bf n} = {\bf k} / k$, one 
recovers the familiar expression for the free-space vacuum
correlation tensor that appears in Eq.(\ref{eqnblochvac}).

Due to translational invariance of the vacuum field,
the free-space correlation tensor only depends on the 
difference vector ${\bf s} \equiv {\bf r}_2 - {\bf r}_1$. 
From rotational invariance, it follows that
the tensor may be decomposed into a scalar part, 
proportional to the unit tensor,
and a quadrupolar part with zero trace, proportional to
$s^i s^j - \frac{ 1 }{ 3 } s^2 \delta^{i,j}$. 
More explicitly, the correlation tensor may be written
for two neighboring points 
\begin{eqnarray}
C^{i,j}_{\infty}( {\bf s})
& = & C^{i,j}_{\infty,0}( {\bf s}) + 
C^{i,j}_{\infty,2}( {\bf s})
	\nonumber\\
& \approx & \big(
1 - {\textstyle \frac{ 7 }{ 30 } } k^2 {\bf s}^2 \big) \delta^{i,j} -
{\textstyle \frac{ 1 }{ 10 } } k^2 
\big( s^i s^j - {\textstyle \frac{ 1 }{ 3 } } {\bf s}^2 \delta^{i,j} \big)
	\label{eq:decomposer-vide}
\end{eqnarray}
We have limited ourselves to second order in $k s$, which
is sufficient to compute the radiation pressure force
and the momentum diffusion tensor since the latter
involve at most a second derivative of the correlation tensor 
[{\em cf.}\ Eqs.(\ref{eq:feed-force-general}, \ref{eq:depart-force-general},
\ref{eq:diffusion-general})].

\subsubsection{Interface contribution}

The interface-dependent part $C^{i,j}_{int}$ of the field correlations
is due to two contributions: 
the evanescent modes ${\bf f}_{\uparrow}({\bf k}_{0\uparrow}, \mu; {\bf r})$
(Eq.(\ref{eq:up-mode}) with $1 < u < n_0$), 
and the crossed term between waves 
incident from above and reflected at the interface
(the two terms of ${\bf f}_{\downarrow}({\bf k}_\downarrow, \mu; {\bf r})$
in Eq.(\ref{eq:down-mode})). Collecting these contributions,
one has
\begin{eqnarray}
&& C^{i,j}_{int}( {\bf r}_1, {\bf r}_2 ) =
	\nonumber\\
&& = \frac{ 3 }{ 8\pi }
\int\limits_{ 1 }^{ n_0 } \! du
\int\limits_{ 0 }^{ 2\pi } \! d\varphi
\frac{ u }{ \sqrt{ n_0^2 - u^2 } }
\sum_\mu
\nonumber\\
&& \quad
e_{\uparrow i}(u, \varphi, \mu)
e_{\uparrow j}^*(u, \varphi, \mu)
|t(u, \mu)|^2
\times \nonumber\\
&& \quad \times
\exp[ i {\bf k}_\Vert ( {\bf r}_{\Vert,1} - {\bf r}_{\Vert,2} )
+  i k_z ( z_1 + z_2 )]
	\nonumber\\
& & + \, 
\frac{ 3 }{ 8\pi }
\int\limits_{ 0 }^{ 1 } \! du
\int\limits_{ 0 }^{ 2\pi } \! d\varphi
\frac{ u }{ \sqrt{ 1 - u^2 } }
\sum_\mu
\nonumber\\
&& \quad
\Big(
e_{\downarrow i}(u, \varphi, \mu)
e_{\downarrow j}^{(r)*}(u, \varphi, \mu)
r^*(u, \mu) 
\, e^{ - i k_z ( z_1 + z_2 )}
\,+
	\nonumber\\
&& \quad
+\,
e_{\downarrow i}^{(r)}(u, \varphi, \mu)
e_{\downarrow j}^{*}(u, \varphi, \mu)
r(u, \mu) 
\, e^{ i k_z ( z_1 + z_2 )}
\Big) 
\times
\nonumber\\
&& \times
\exp[ i {\bf k}_\Vert ( {\bf r}_{\Vert,1} - {\bf r}_{\Vert,2} )]
	\label{eq:corr-int-u-du}
\end{eqnarray}
(Recall that ${\rm Im}\, k_z > 0$ in the first integral
and $k_z > 0$ in the second.)
It is now evident that the interface contribution
only depends on the in-plane
difference vector ${\bf s}_\Vert = {\bf r}_{\Vert,2} 
- {\bf r}_{\Vert,1}$ and the sum of the distances $z_1 + z_2$. 

We perform the integration over the azimuthal angle $\varphi$
with the help of the following formula \cite{Abramowitz}
and its derivatives with respect to $k s_\Vert$
\begin{equation}
\int\limits_0^{2\pi} \! d\varphi \,
e^{ - i k s_\Vert u \cos\varphi } = 2 \pi J_0( k s_\Vert u )
\end{equation}
where $J_0$ is the Bessel function of zeroth order.
We also observe the following property of the Fresnel coefficients
\begin{equation}
n_0 > u > 1: \quad
\sqrt{ \frac{ u^2 - 1 }{ n_0^2 - u^2 } } 
| t( u, \mu ) |^2
= 2 \, \text{Im} \, r( u, \mu )
	\label{eq:relation-rt}
\end{equation}
that allows to combine the contributions of reflected
and evanescent modes in a compact way.
Finally, the correlation tensor is decomposed into its isotropic,
axial and quadrupolar parts, according to
\begin{mathletters}
	\label{eq:decomposer-int}
\begin{eqnarray}
&& C^{i,j}_{int}( z ; {\bf s}_\Vert) =
C^{i,j}_{int}( {\bf r} - {\textstyle \frac12} {\bf s},
{\bf r} + {\textstyle \frac12} {\bf s})
	\nonumber\\
&& = C^{i,j}_{int,0}( z ; {\bf s}_\Vert)
+ C^{i,j}_{int,1}( z ; {\bf s}_\Vert)
+ C^{i,j}_{int,2}( z ; {\bf s}_\Vert)
	\label{eq:corr-tenseur}\\
&& C^{i,j}_{int,0}( z ; {\bf s}_\Vert) =
c_0 (z; {\bf s}_\Vert^2) 
\delta^{i,j}
	\label{eq:scalar-int}\\
&& C^{i,j}_{int,1}( z ; {\bf s}_\Vert) =  
k a( z; {\bf s}_\Vert^2 ) \Big(
\delta^{z,i} s_{\Vert}^{j} - s_{\Vert}^{i} \delta^{z,j} \Big)
	\label{eq:axial-int}\\
&& C^{i,j}_{int,2}( z ; {\bf s}_\Vert) =
q_0( z; {\bf s}_\Vert^2 )
\Big( \delta^{i,z} \delta^{j,z} 
- {\textstyle \frac{ 1 }{ 3 } } \delta^{i,j} \Big) 
+ \nonumber\\
&& \quad
+ k^2 q_2( z; {\bf s}_\Vert^2 )
\Big( s_{\Vert}^{i} s_{\Vert}^{j}
- {\textstyle \frac{ 1 }{ 2 } } {\bf s}_\Vert^2  
( \delta^{i,j} - \delta^{i,z} \delta^{j,z} ) \Big) 
	\label{eq:quadrupolar-int}
\end{eqnarray}
In these expressions, the dimensionless weight functions 
$c_0, q_0, a, q_2$ are given by the Sommerfeld integrals
\end{mathletters}
\begin{mathletters}
	\label{eq:def-c0etc}
\begin{eqnarray}
c_0 (z; {\bf s}_\Vert^2) & = & \frac 12 \,\text{Re}
\int_0^{n_0} \! \frac{ du \, u }{ v } \, J_0( k s_\Vert u )
\times \label{eq:def-c0}\\
&& \times
( r_{TE} + ( 2 u^2 - 1 ) r_{TM} )
\exp{ 2 i k z v }
	\nonumber\\	
q_0( z; {\bf s}_\Vert^2 ) & = & \frac 34 \,\text{Re}
\int_0^{n_0} \! \frac{ du \, u }{ v } \, J_0( k s_\Vert u ) 
\times \label{eq:def-q0}\\
&& \times 
( - r_{TE} + ( u^2 + 1 ) r_{TM} )
\exp{ 2 i k z v }
	\nonumber\\
a( z; {\bf s}_\Vert^2 ) & = & \frac 32 \,\text{Im}
\int_0^{n_0} \! du \, u^2 \, 
\frac{ J_1( k s_\Vert u ) }{ k s_\Vert } 
r_{TM}\exp{ 2 i k z v }
	\label{eq:def-a}\\
q_2( z; 0 ) & = & \frac{ 3 }{ 2 } \,\text{Re}
\int_0^{n_0} \! \frac{ du \, u }{ v }
\frac{ J_2( k s_\Vert u ) }{ (k s_\Vert)^2 }
\times \label{eq:def-q2}\\
&& \times
( r_{TE} - ( u^2 - 1 ) r_{TM} ) \exp{ 2 i k z v }
	\nonumber
\end{eqnarray}
These integrals are computed numerically and are plotted 
as a function of $kz$ in Fig.~\ref{fig:c0etc} 
for $s_\Vert = 0$. The refractive index is $n_0 = 1.5$.
\end{mathletters}

\subsection{Relation to field susceptibility}
\label{a:susceptibility}

As a final comment, we would like to display the link
between the vacuum correlation tensor on the one hand,
and the classical field susceptibility, on the other.
The latter quantity gives the (positive-frequency)
electric field created at position ${\bf r}_1$ 
by an oscillating dipole located at ${\bf r}_2$:
\begin{equation}
E_i( {\bf r}_1 ) e^{- i \omega t} 
= G^{i,j}( {\bf r}_1, {\bf r}_2 ) D_j e^{- i \omega t}.
\end{equation}
The tensor $G^{i,j}$ may be calculated from classical
electrodynamics \cite{Jackson,Maradudin75,Agarwal75a}.
For the vacuum--dielectric interface, we have checked
the following relation to the vacuum correlation tensor
\begin{equation}
{\cal E}^{i,j}( {\bf r}_1, {\bf r}_2 ) =
2 \hbar \,
{\rm Im}\, G^{i,j}( {\bf r}_1, {\bf r}_2 ) .
\label{eq:coincidence}
\end{equation}
This result shows that, at least for zero temperature
and a two-level system, both the spontaneous emission rates
and the associated forces may be calculated from a
classical field calculation alone, without explicitly
quantizing the field. In particular, Eq.(\ref{eq:coincidence})
justifies our interpretation of the radiation pressure force
in terms of a classical picture where the atomic dipole
interacts with its own radiation reaction field 
reflected from the vacuum--dielectric interface.

% end of app-c.tex

% -------------------------------------------------------
% app-d.tex
% -------------------------------------------------------

\section{Optical pumping in a $TM$-polarized evanescent wave}
\label{a:pompage-TM}

In this appendix, we study the optical pumping of a $J=1/2$ atom
in an evanescent wave with $TM$ polarization. The helicity 
(\ref{eq:depl-lum-TM}) is parallel to the $y$-axis that we choose
as the quantization axis. The field is then a superposition of $\sigma^+$
and $\sigma^-$ circular components, and the light-shift operator~%
(\ref{eq:precession}) is diagonal. It may also
be verified from the optical pumping equation Eq.(\ref{eq:eq-pompage-tenseurs})
that the magnetization components $J_{x,z}$ decouple from $J_{y}$ 
and the total population. 
Assuming that the atoms are initially unpolarized, 
the pumping process only depends on
$J_{y}$ whose evolution is given by
(we put again $w = 1$ in the optical pumping equation) 
\begin{mathletters}
\label{eq:pompage-TM}
\begin{eqnarray}
\left. \frac{\partial J_{y}}{\partial t}\right| _{0} &=&
- \Gamma_p( z ) - \Gamma_{y,relax}( z ) J_y 
\\
\Gamma_p( z ) & = &
2 \alpha ^{2}\Gamma'_\infty e^{-2\kappa z} 
c_\Vert( z ) \frac{ 2 \kappa Q }{ k^2 }
\label{eq:taux-pompage-TM}
\\
\Gamma_{y,relax}( z ) & = &
2\alpha^2 \Gamma'_\infty e^{-2\kappa z} 
c_\Vert( z ) \frac{ \kappa^2 + Q^2 }{ k^2 }
\label{eq:taux-relax-y}
\end{eqnarray}
Optical pumping hence builds up a net magnetization 
along the negative $y$ axis
with a rate $\Gamma_p( z )$ that only depends on the fluorescence rate 
$\Gamma_\infty c_\Vert( z )$
for a polarization parallel to the interface.
This may be understood from the rate equations for the populations 
$w_\pm = \frac12 ( 1\pm J_{y} )$ of the Zeeman sublevels 
$|\pm 1/2\rangle _{y}$ with respect to the $y$ axis. 
From Eqs.(\ref{eq:pompage-TM}), the following rate equations are easily
found 
\end{mathletters}
\begin{mathletters}
\label{eq:eq-pompage-taux-TM}
\begin{eqnarray}
\frac{\partial w_{+}}{\partial t} &=&
-\Gamma_{+\to-}( z ) w_+
+\Gamma_{-\to+}( z ) w_-
\\
\frac{\partial w_{-}}{\partial t} &=&
-\Gamma_{-\to+}( z ) w_-
+\Gamma_{+\to-}( z ) w_+
\\
\Gamma_{-\to+}( z ) &=&
{\textstyle \frac 12} [ \Gamma_{y,relax}( z ) - \Gamma_p( z ) ]
\nonumber\\
& = &
2 \alpha ^{2}\Gamma _{\infty }^{\prime }\,e^{-2\kappa
z}c_{\Vert }(z)\frac{(\kappa -Q)^{2}}{2 k^{2}}
\label{eq:taux-minus->plus}
\\
\Gamma_{+\to-}( z ) &=&
{\textstyle \frac 12} [ \Gamma_{y,relax}( z ) + \Gamma_p( z ) ]
\nonumber\\
& = &
2 \alpha ^{2}\Gamma _{\infty }^{\prime }\,e^{-2\kappa
z}c_{\Vert }(z)\frac{(\kappa + Q)^{2}}{2 k^{2}}
\label{eq:taux-plus->minus}
\end{eqnarray}
As expected, the rate for the transition 
$|-1/2\rangle _{y}\to |+1/2\rangle _{y}$, {\em e.g.}, is proportional
to the square $|\xi_{0+}|^2 = |(\kappa - Q)/\sqrt{2}k|^2$ 
of the amplitude of the $\sigma ^{+}$ polarization 
in the evanescent field (the
circular polarization being defined with respect to the $y$ axis).  
In this process, the atom absorbs a 
$\sigma _{+}$ polarized photon from the driving field
and spontaneously emits a $\pi $ polarized photon
({\em cf.}\ Fig.~\ref{fig:cycles-fluo}c).
The electric field of the latter being parallel to
the $y$ axis, the spontaneous transition rate is proportional to 
the coefficient $C^{y,y}( z ) = c_\Vert( z )$ of the vacuum
correlation tensor. 

The opposite transition
$|+1/2\rangle _{y}\to |-1/2\rangle _{y}$ has a larger rate~%
(\ref{eq:taux-plus->minus}) because in the $TM$ case,
the amplitude $%
\xi _{0-}=(\kappa +Q)/\sqrt{2}k$ of the $\sigma ^{-}$ polarization is
stronger than $\xi _{0+}$.
The steady-state magnetization is not maximum, however, 
because the pumping rate $\Gamma_p$ (\ref{eq:taux-pompage-TM})
is smaller than the relaxation rate
$\Gamma_{y,relax}$~(\ref{eq:taux-relax-y}),
the polarization of the $TM$ evanescent wave being not purely circular. 
The steady-state solution of the Eq.(\ref{eq:pompage-TM}) yields
\end{mathletters}
\begin{eqnarray}
&& w_{-}^{(stat)}-w_{+}^{(stat)}
= - J_{y}^{(stat)} 
= \nonumber\\
&& = \frac{ \Gamma_p( z ) }{ \Gamma_{y,relax}( z ) }
= \frac{2\kappa Q}{\kappa ^{2}+Q^{2}} < 1 ,
\label{eq:Jy-stat-TM}
\end{eqnarray}
as stated in the text.

\end{document}